\documentclass[10pt,journal,compsoc]{IEEEtran}
\ifCLASSOPTIONcompsoc
  \usepackage[nocompress]{cite}
\else
  \usepackage{cite}
\fi
\ifCLASSINFOpdf
  \usepackage[pdftex]{graphicx}
\else
\fi
\usepackage{amsmath}
\DeclareMathOperator*{\argmax}{argmax}

\usepackage{algorithm}
\usepackage{algorithmic}
\usepackage{caption}
\usepackage{array}

\usepackage{multirow}
\usepackage{makecell}
\usepackage{tabularx,booktabs}
\newcolumntype{C}{>{\centering\arraybackslash}X} 
\usepackage{lipsum} 

\ifCLASSOPTIONcompsoc
  \usepackage[caption=false,font=footnotesize,labelfont=sf,textfont=sf]{subfig}
\else
 \usepackage[caption=false,font=footnotesize]{subfig}
\fi
\usepackage{fixltx2e}

\usepackage{stfloats}
\begin{document}
%
\title{PRISM: A Hierarchical Intrusion Detection Architecture for Large-Scale Cyber Networks}
%
%
%
%

\author{Yahya~Javed,
        Mosab~A.~Khayat,
        Ali~A.~Elghariani,
        and~Arif~Ghafoor 
\IEEEcompsocitemizethanks{

\IEEEcompsocthanksitem Yahya Javed and Arif Ghafoor are with the Elmore Family School of Electrical and Computer Engineering, Purdue University, West Lafayette, IN, USA.\protect\\
E-mail: yjaved@purdue.edu, ghafoor@ecn.purdue.edu

\IEEEcompsocthanksitem Mosab A. Khayat is with the Department of Computer Engineering, Umm Al-Qura University, Makkah, KSA. \protect\\
Email: maakhayat@uqu.edu.sa

\IEEEcompsocthanksitem Ali A. Elghariani is with XCOM-Labs, San Diego, CA, USA. \protect\\
Email: ali.elghariani@gmail.com




}


}

\IEEEtitleabstractindextext{%
\begin{abstract}
The increase in scale of cyber networks and the rise in sophistication of cyber-attacks have introduced several challenges in intrusion detection. The primary challenge is the requirement to detect complex multi-stage attacks in realtime by processing the immense amount of traffic produced by present-day networks. In this paper we present PRISM, a hierarchical intrusion detection architecture that uses a novel attacker behavior model-based sampling technique to minimize the realtime traffic processing overhead. PRISM has a unique multi-layered architecture that monitors network traffic distributedly to provide efficiency in processing and modularity in design. PRISM employs a Hidden Markov Model-based prediction mechanism to identify multi-stage attacks and ascertain the attack progression for a proactive response. Furthermore, PRISM introduces a stream management procedure that rectifies the issue of alert reordering when collected from distributed alert reporting systems. To evaluate the performance of PRISM, multiple metrics have been proposed, and various experiments have been conducted on a multi-stage attack dataset. The results exhibit up to 7.5x improvement in processing overhead as compared to a standard centralized IDS without the loss of prediction accuracy while demonstrating the ability to predict different attack stages promptly.  
\end{abstract}

}

\maketitle

\IEEEdisplaynontitleabstractindextext

%
\IEEEpeerreviewmaketitle

\IEEEraisesectionheading{\section{Introduction}\label{sec:introduction}}

%
%
%
%
\IEEEPARstart{T}{he} past couple of years have witnessed a concerning proliferation of cyber-attack incidents with the targets ranging from critical government infrastructures to common internet users. State-sponsored Advanced Persistent Threats (APTs) have revealed the limitations of current Intrusion Detection Systems (IDSs) with median dwell time of 30 days and internal detection rate on a continuous decline for the past couple of years \cite{fireEye2020}. Today, IDSs face three major challenges; an ever increasing attack complexity, astronomical increase in data traversing through the networks, and the need to detect attacks in realtime with in-depth information. To account for the growing attack complexity, modern IDSs evolved from simple string matching systems to sophisticated machine learning-based frameworks, making intrusion detection a computation intensive process \cite{lukaseder}. To cater for the processing of massive data, there is no substantial change in the IDS computational process other than the introduction of more powerful processing hardware. However, according to Gilder's law the bandwidth grows at least three times faster than the compute power. While, compute power doubles every eighteen months (Moore's law), bandwidth doubles every six months \cite{gilder}. This widening gap between the computational capacity and network bandwidth has made it challenging for IDSs to process every packet in realtime and identify threats in a cost effective manner. Consequently, as the limited IDS resources are deployed at only few parts of the network, the chances of attackers penetrating the network without detection increase. At the same time, it is expected of the IDSs to identify threats with actionable information that can lead to a proactive response, contrary to binary classification like alerts that have little use for response operations\cite{sommer}. These challenges necessitate an overhaul in the IDS architecture and the intrusion detection process.

\par Commercial and opensource IDSs for enterprise networks are generally deployed in a centralized orchestration. Some of these IDSs use multiple parallel processing nodes to handle large quantity of data \cite{zeekCluster}. The network traffic is captured by taping the link connecting the enterprise network to the internet which is then processed by the IDS detection module to find malicious packets \cite{snort, zeek}. This design choice has several disadvantages including high cost of hardware that can process data in Gbps of data rates, limitations in scalability as the network grows over time, and inability to monitor internal network traffic that ignores attack sources other than the internet. In literature, there is a significant amount of work that proposes a distributed IDS design. While the proposed distributed IDS architectures address the problems associated with a centralized architecture, they introduce new issues in addition to ignoring some of the aforementioned challenges experienced by the IDSs. As discussed in the related work section of the paper, some of the proposed distributed IDS architectures only focus on enhancing the traffic processing capabilities, but at the same time have the inability to detect complex multi-stage attacks, and do not possess intrusion prediction feature that can provide the response mechanism with enough information for a proactive response. On the other hand, the proposed IDSs with intrusion prediction capabilities present a conceptual distributed architecture that does not address the practical challenges associated with a distributed design. Among those challenges is the alert reordering issue that occurs when alerts are collected from distributed sources with different network latencies. This can cause prediction error in time series data-based prediction models that are generally employed to detect multi-stage attacks. Additionally, the proposed IDSs with intrusion prediction capabilities lack in proper implementation and evaluation in a distributed alert reporting setup. This entails the need to develop an intrusion detection architecture that can process network traffic in parallel, predict the attack progression and detect complex multi-stage attacks, while addressing the challenges introduced by the distributed design.           

\par Present-day IDSs process every packet traversing through the network in their assigned region of observation. For each packet or flow, a decision is made if they are malicious or benign. Most of the traffic in the network is benign, and even during an attack, the malicious traffic is a fraction of the total network traffic. Moreover, once an attack is detected, most of the malicious packets that follow are of the same continuing attack. These observations exhibit that there is room for optimizations in the way IDSs monitor traffic and detect threats. Network traffic sampling is a plausible approach to make intrusion detection much more efficient by directing only sampled traffic to the IDS. However, existing network sampling techniques are designed for network traffic engineering purposes and perform poorly for intrusion detection applications \cite{mai}. This necessitates the development of a network intrusion detection centric sampling scheme with the objectives of providing acceptable IDS accuracy, retention of IDS's ability to identify every attack and maximize attack detection speed by reducing the amount of traffic to be processed by the IDS.       

\par In this paper we introduce PRISM: Performance-oriented Realtime Intrusion-detection and Security Monitoring. The fundamental design goals of PRISM are to detect complex multi-stage attacks in high bandwidth cyber networks with minimum processing overhead, provide attack progression prediction and holistic security assessment of the entire network in realtime, and ensure scalability by comprising of a modular architectural construct. The distinct contributions of PRISM are as follows:

\par \emph{1) Intrusion detection, prediction and security analysis:} PRISM incorporates a multi-layered structure that utilizes the services of its several internal systems to detect intrusions, predict attack progression and present a holistic security outlook of the system. The network traffic is monitored in a distributed fashion by dividing the network into different surveillance zones. Each surveillance zone is supervised by an IDS that reports its local alerts. The alerts are correlated in a Hidden Markov Model (HMM)-based prediction system that first determines the type of multi-stage attack and then predicts its current stage. The prediction output and alerts from surveillance zones are then processed to compute and visualize several metrics for intrusion response facilitation.


\par \emph{2) Threat-aware sampling:} PRISM uses a probabilistic sampling scheme to process more traffic from devices that have more likelihood of being attacked. The sampling scheme utilizes a novel attacker behavior model-based ranking mechanism that ranks devices in the network according to their vulnerability and inter-device reachability information. Packets are sampled according to a sampling probability which is calculated using the rank-based score of the packet's source and destination device. Instead of processing every captured packet, only sampled packets are forwarded to the detection module of the IDS. Using the attacker behavior model-based sampling, PRISM strategically focuses on the traffic from those devices that are critical, and are more prone to be attacked, reducing processing overhead and the required computing resources.

\par \emph{3) Alert stream management:} PRISM manages the alert streams from multiple surveillance zones by implementing an efficient alert stream management mechanism. When alerts are received from several distributed sources, they can be at a different order in contrast to the order in which they are generated. The principal reason for this reordering is the transmission of alerts from different surveillance zones using different network connections with varying latencies. The significance of alert stream management mechanism is that it restores the alert order in realtime which otherwise would result in prediction error when evaluated by the prediction system.


\par \emph{4) Experimentation and performance evaluation:} The performance of PRISM is evaluated through extensive experimentation on ISCX-2012 \cite{iscx} and CIC-2017 \cite{cic} datasets. The datasets consist of real traffic capture of several complex multi-stage attacks launched on their respective target networks for multiple days. Experiments are designed to test the performance of all internal systems of PRISM in terms of multiple performance metrics. Results show that PRISM can maintain good prediction accuracy while significantly reducing the processing overhead in addition to being able predict attack progress notably early when studied under several experimental conditions. The security analysis capability of PRISM provides valuable utility by computing and visualizing cyber situational awareness-centric metrics.   

\par The rest of this paper is structured as follows. Section 2 presents the relevant background. Section 3 explains the working of PRISM and its constituent systems. The experimental setup is described in Section 4. The performance evaluation of PRISM is provided in section 5. The related work is outlined in Section 6. Finally, Section 7 concludes the paper.


 




\section{Background}
PRISM is built by utilizing the concepts of attacker behavior modeling, HMM and stream processing. In this section, the basics of the underlying methods and models used in PRISM are discussed.

\subsection{Attacker Behavior Modeling}
Attacker behavior modeling is a widely used technique to evaluate the security of enterprise networks \cite{jonsson}. PRISM models the attacker behavior by extending the Google PageRank mechanism. PageRank is flexible enough to model a diverse set of cases and its employment in network security related graph problems is not uncommon \cite{mehta, sawilla}. The behavior of a random web-surfer is modeled by PageRank who starts at a web-page and keeps on clicking links until gets bored and starts at another web-page. A web-page has higher rank if many pages link to it, and a page is important if it has a high rank and has few links to other pages. To model the probability of the web-surfer breaking the chain of clicking links and starting from a new random web-page, a damping factor is introduced which essentially asserts that long chains of page clicking is unlikely \cite{google}. PageRank iteratively calculates the rank of each page using Eq. \ref{pagerank}.

\begin{equation}
    PR(U) = 1-\delta + \delta \sum_{j \in In(U)} \frac{PR(W_j)}{Out|W_j|}
    \label{pagerank}
\end{equation}

Where, $PR(U)$ is the PageRank value of a page $U$, $\delta$ is the damping factor with the value commonly set to 0.85, $PR(W_j)$ is the PageRank value of page $W_j$ that points to page $U$, and $Out|W_j|$ is the number of links going out of page $W_j$. 

\subsection{Hidden Markov Model}
Markov chains are used to determine the probability of a sequence of events that are observable. However, there are many instances in which the events of interest are hidden. A discrete first-order HMM is a doubly stochastic model that accommodates both observed and hidden events. An HMM is characterized by the following elements \cite{HMMRabiner}.

\par 1) The set of N states represented by $S = \{s_1, s_2, .., s_N\}$, where the state at time $t$ is $q_t$.
\par 2) A set of M distinct observation symbols per state denoted as $V = \{v_1, v_2, .., v_M\}$.
\par 3) Transition probability matrix $A_{NxN}$ where $a_{ij}$ is the probability of shifting from state $s_i$ to state $s_j$.
\begin{equation}
    a_{ij} = P(q_{t+1} = s_j|q_t = s_i), \;\;\;\;\;\;\; i,j \in [1, N] 
    \label{Amatrix}
\end{equation}
\par 4) Emission probability matrix $B_{NxM}$, where $b_i(k)$ is the probability of an observation $v_k$ being generated in a state $s_i$.
\begin{equation}
\begin{split}
    b_i(k) = P(v_k \; at \; t|q_t = s_i), \;\;\;\;\;\;\; & i\in [1, N] \\ 
    & k \in [1, M]    
\end{split}
\label{Bmatrix}
\end{equation}
\par 5) Initial probability vector $\pi=\{\pi_1, \pi_2, .., \pi_N\}$, where $\pi_i$ is the probability of Markov chain initializing at state $s_i$.
\begin{equation}
    \pi_{i} = P(q_1 = s_i), \;\;\;\;\;\;\; i\in [1, N]
    \label{Pimatrix}
\end{equation}




\par HMM-driven prediction systems use the model $\lambda = (S, V, A, B, \pi)$ and observation sequence $O = \{o_1, o_2, .., o_T\}$ to effectuate various tasks. Note that observations are IDS alerts in our case and we use the terms observation sequence and alert sequence interchangeably. The use of HMMs in practical applications is characterized by three fundamental problems. First, how to determine $P(O|\lambda)$, the probability of observation sequence given the model. Second, how to discover the best state sequence $Q = {q_1, q_2, .., q_T}$ given the observation sequence and model. Third, how to learn the model parameters $A$ and $B$ that maximize $P(O|\lambda)$ given the observation sequence and the set of states. The first problem is solved using the forward algorithm, the second problem is solved using the Viterbi algorithm and for the third problem the standard solution is unsupervised learning-based Baum-Welch algorithm, but in our implementation we follow a supervised learning approach discussed in Section 3 of the paper.

\subsection{Stream Processing}
In the event of an attack, a continuous stream of alerts is generated by the IDSs. PRISM employs stream processing techniques to perform certain data manipulation operations on this stream of alerts. In the stream processing context, the terms bounded and unbounded refers to the finite and infinite data, respectively. Performing data processing operations like sort, aggregation etc. on unbounded data causes semantic problems, therefore, it is important to have some notion of bounds on the unbounded data. Windowing is a process of dividing data into finite blocks that can be processed as a group. The process of windowing is essential to analyze unbounded data and it has been proven to be useful for bounded data as well in certain scenarios. Windowing is generally time-based or count-based, however, there are several other implementations as well. As the names suggest, a time-based window defines a data block on the data stream with respect to a time period, while grouping a certain number of data items together forms a count-based window. The three main window types are tumbling, sliding and sessions. Tumbling windows are non-overlapping windows defined by constant window size and a data item cannot be the member of more than one window. Sliding windows are defined by a fixed window size and a slide period which may be less than the window size implying that the windows can overlap. In a sliding-windows-based implementation, a data item can be part of multiple windows. Sessions are windows that are defined over a subset of data using a timeout gap. The events that occur within a timeout are combined to form a session \cite{googleDataFlow}. PRISM uses a custom implementation of the windowing process employing both time-based and count-based windows to manage the flow of alerts streams.


\section{System Model and Architecture}
\begin{figure}[!t]
\centering
\includegraphics[width=3in,height=4in]{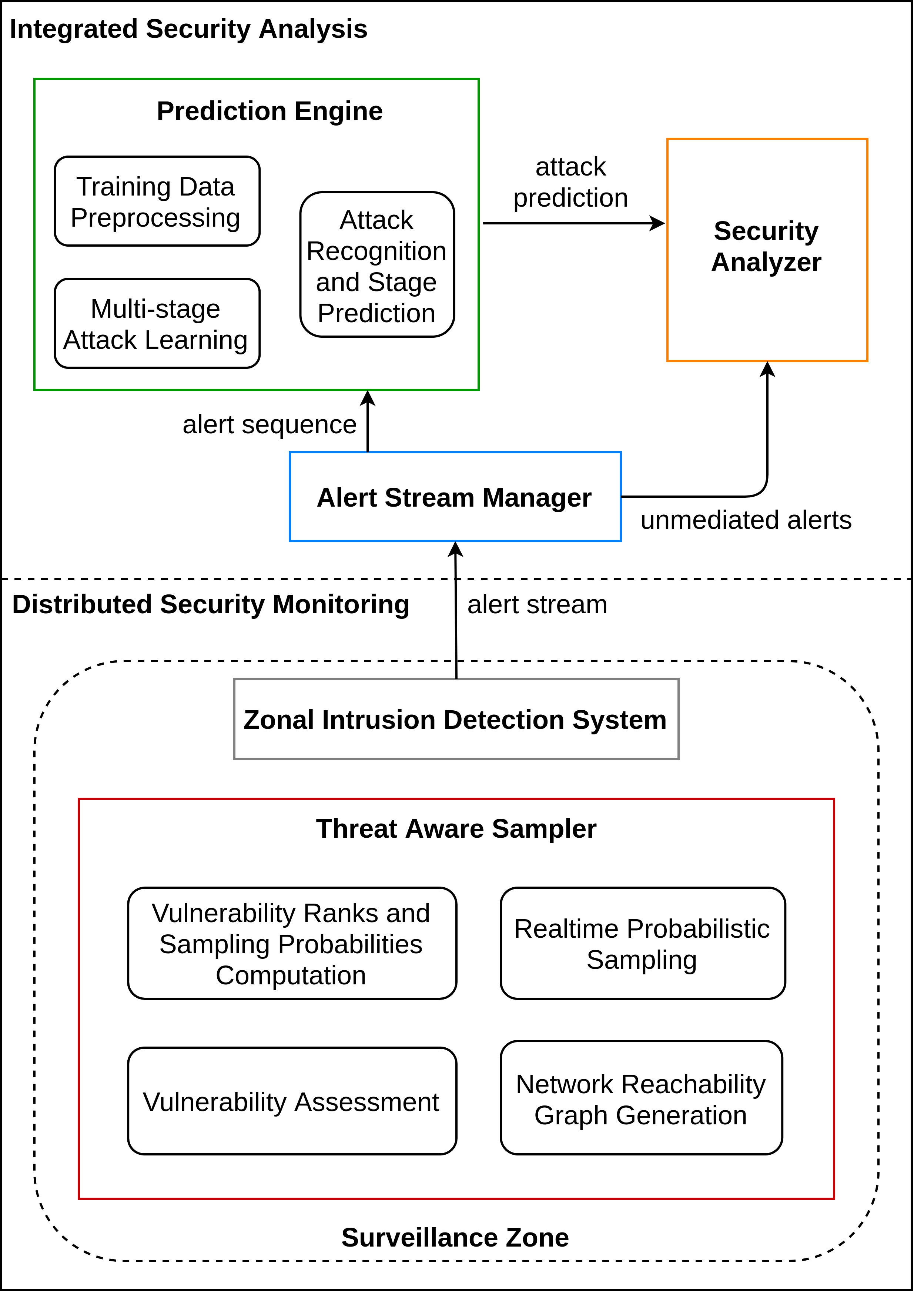}
\caption{PRISM and its constituent systems.}
\label{arch1}
\end{figure}

PRISM is depicted in Fig. \ref{arch1} along with its five constituent systems: Threat-Aware Sampler (TAS), Zonal IDS, Alert Stream Manager (ASM), Prediction Engine, and Security Analyzer. TAS and Zonal IDS implement the distributed security monitoring functionality of PRISM and are replicated among all surveillance zones. In Fig. \ref{arch1} only one surveillance zone is shown for lucidity. The demarcation of surveillance zones can be done based on the network layout, traffic volume distribution and amount of available zonal IDS resources. The objective is to distribute the traffic equally among surveillance zones with the consideration of network administrative constraints. The number of available zonal IDS resources determine the number of surveillance zones. ASM, Prediction Engine and Security Analyzer implement the integrated security analysis functionality of PRISM. The detailed information about the design and working of each component of PRISM is discussed as follows.


\subsection{Threat-Aware Sampler}
TAS is deployed in every surveillance zone along with the zonal IDS. Mirrored traffic from the router of the surveillance zone is forwarded to the TAS which samples traffic in realtime and directs it to the zonal IDS. TAS has offline and online modes of operation. In offline mode, TAS performs three operations: vulnerability assessment, network reachability graph generation and vulnerability ranks and sampling probabilities computation. In online mode, TAS carries out the task of realtime probabilistic sampling. The functioning of the internal modules of TAS is discussed in detail.

\subsubsection{Vulnerability Assessment}
The vulnerability assessment module of TAS performs vulnerability scan of each device in the network to determine their Common Vulnerability Scoring System (CVSS) scores \cite{cvss}. CVSS is a popular measure of assessing the severity of a device's vulnerabilities by assigning a score out of 10. CVSS scores are computed using the information of three metric groups: base, temporal and environmental. For base metric group, information of five exploitability metrics and three impact metrics is required to calculate the score of a vulnerability. The vulnerability score for a device can be obtained by considering the CVSS scores of all vulnerabilities identified in a device. There are many commercial and opensource tools available that can scan the network and assess the vulnerabilities of each device \cite{nessus, openvas}. TAS has the flexibility to operate with any proprietary or opensource CVSS compatible network vulnerability scanner. The vulnerability assessment procedure adapted for experimentation is discussed in Section 4 of the paper.   

\subsubsection{Network Reachability Graph Generation}
\begin{figure}[!t]
\centering
\includegraphics[width=2.5in]{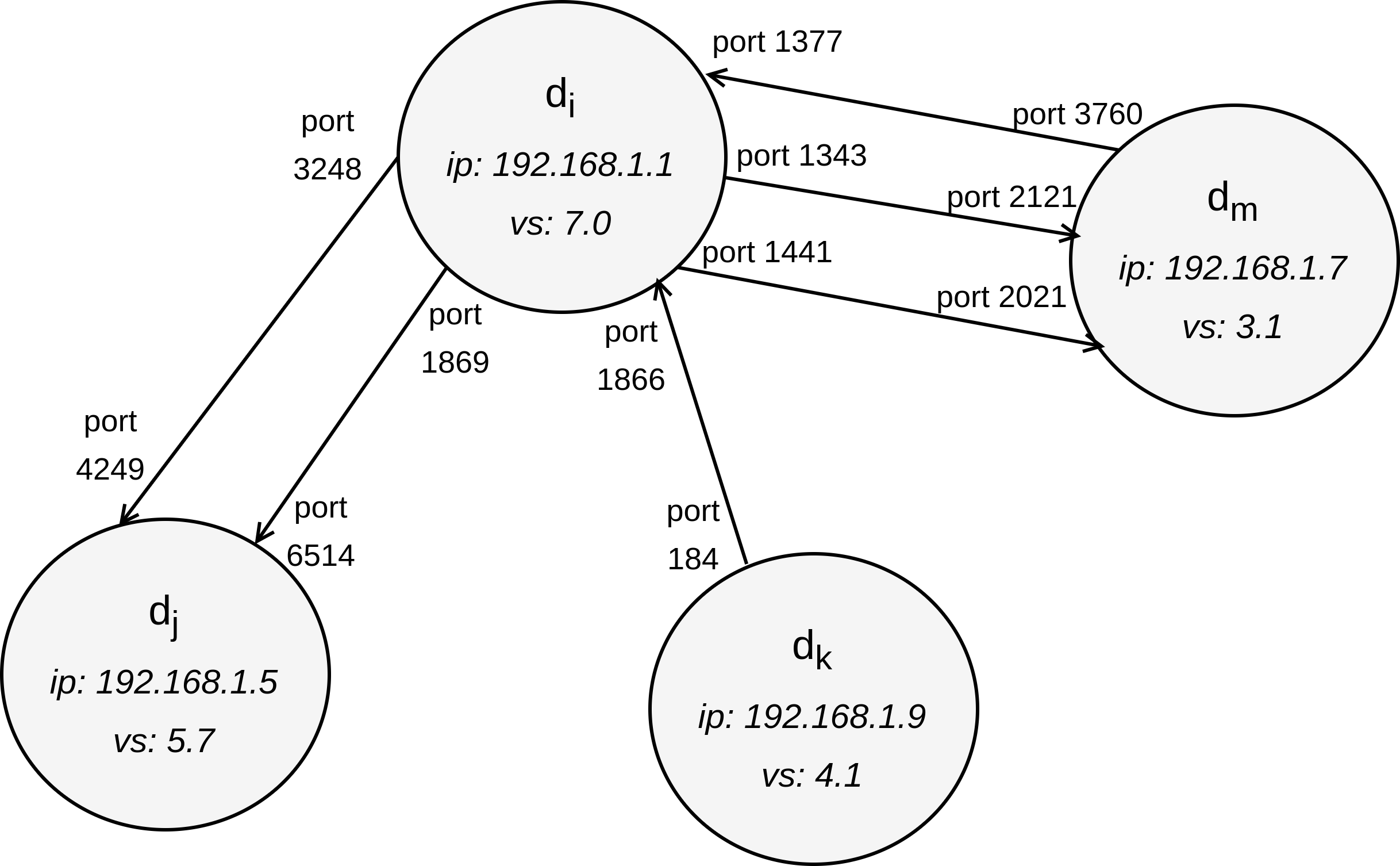}
\caption{A sample 4-device NRG.}
\label{nrg}
\end{figure}

TAS employs the Network Reachability Graph (NRG) to develop the attacker behavior model that is used to rank network devices according to their likelihood of being compromised. NRG$<D,E>$ is a directed mutligraph with its vertices corresponding to the devices in the network represented by $D = \{d_1, d_2, ..., d_n\}$. The parallel edges of NRG correspond to the end-to-end communication between devices on different ports and are represented by $E$, a multiset of edges (ordered pairs of vertices). A device in the network is identified by its $ip$ address, and for each device, the vulnerability score $vs$ computed by the vulnerability assessment module is also maintained. The communication among network devices is in part determined by firewall policies, network configuration and authentication/permissions within devices. A sample NRG is illustrated in Fig. \ref{nrg}. NRG can be generated using any network mapping tool like Nmap \cite{nmap}. Another method of generating NRG is by mapping the flow information of monitored network activity over time. We have employed the latter method that is discussed in Section 4 of the paper. 

\subsubsection{Vulnerability Ranks and Sampling Probabilities Computation}
\begin{figure}[!t]
\centering
\includegraphics[width=2.2in, height = 3in]{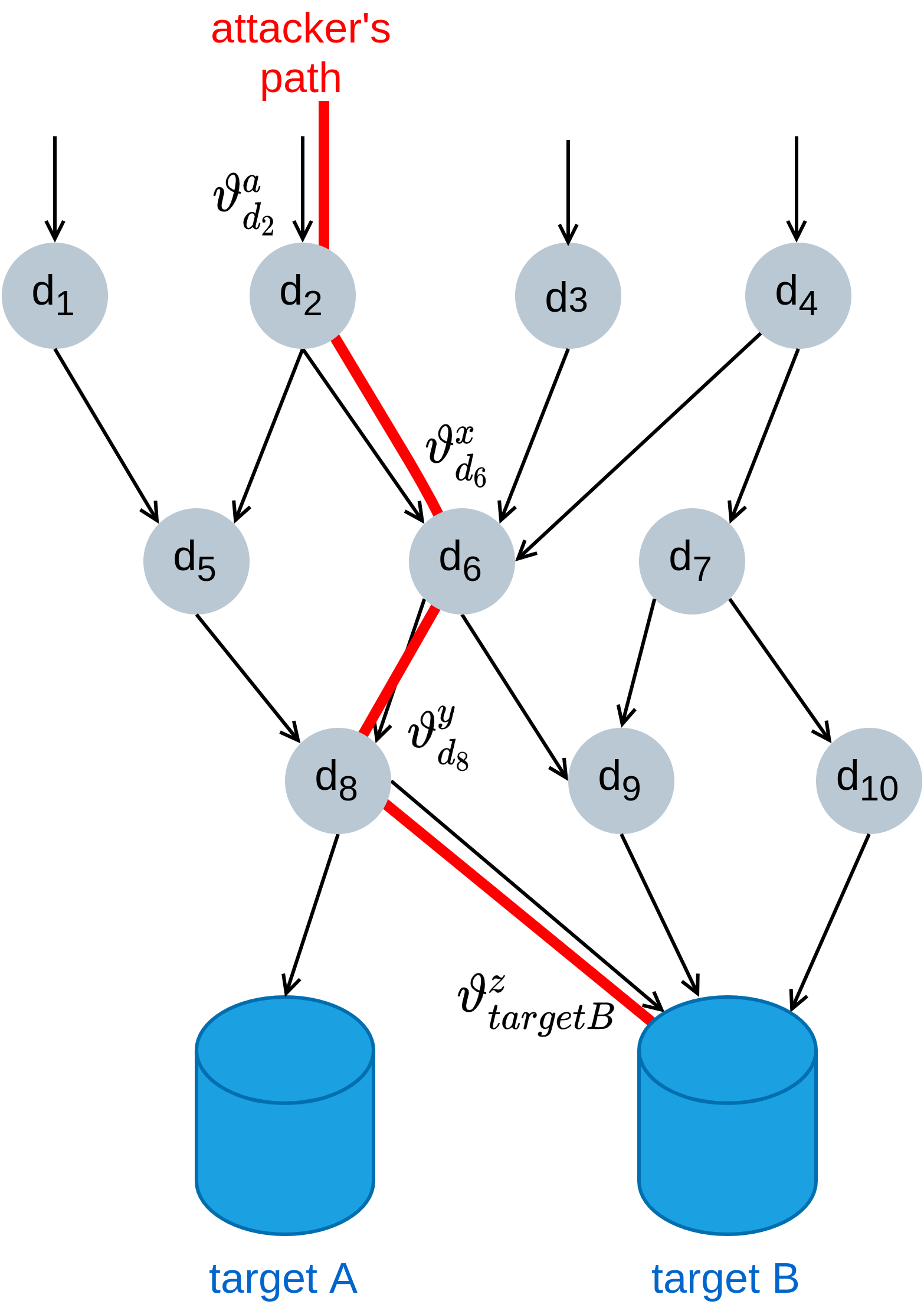}
\caption{Illustration of attacker's movement in the network.}
\label{abehav}
\end{figure}

The core idea behind threat-aware sampling is to intelligently sample traffic by modeling the attacker behavior that can rank network devices according to their likelihood of being targeted by the attacker. The proposed attacker behavior model captures the potential actions of an attacker who has a specific target and tries to penetrate the network through any accessible device. From the compromised devices, the attacker tries to strike every other device until the attainment of target. Fig. \ref{abehav} shows an example attack in which the attacker compromises devices $d_2,$ $d_6,$ and $d_8$ to attack $targetB$ by exploiting vulnerabilities $\vartheta_{d_2}^a$, $\vartheta_{d_6}^x$, $\vartheta_{d_8}^y$ and $\vartheta_{targetB}^z$, respectively. The attacker behavior can be modeled by extending the random surfer model with security-centric semantics. A device has a high vulnerability rank if it has high vulnerability score and has wide attack surface i.e. several devices can communicate with it on multiple ports. A device's vulnerability rank is also high if the devices communicating to it have high vulnerability scores. The damping factor in the attacker behavior model corresponds to the fact that it is unlikely for an attacker to take a longer path to reach target if a shorter path exists. The extended PageRank formalism that captures the vulnerability transference effect of inter connected devices is presented in Eq. (\ref{rankeq}). Where, $\tau_i$ represents the vulnerability transference effect of a device $d_i$, $\delta$ is the damping factor, $In(d_i)$ is the in-set of device $d_i$ i.e. it is the set of devices transmitting to $d_i$, $\rho_j$ is the vulnerability rank of device $d_j$ which is one of the devices in the in-set of $d_i$, $|Out(d_j)|$ is cardinality of the set of outgoing edges from device $d_j$, and $|Out(d_j, d_i)|$ is cardinality of the set of edges from device $d_j$ to $d_i$.   

\begin{equation}
    \tau_i = 1-\delta + \delta \sum_{j \in In(d_i)} \rho_j \frac{|Out(d_j, d_i)|}{|Out(d_j)|}
    \label{rankeq}
\end{equation}


\par A device receives its vulnerability transference effect from the devices that can communicate with it and the devices that can communicate using multiple connections (port pairs) contribute more to the vulnerability transference. In vulnerability rank computation of a device, the device's self vulnerability contributes to its rank in addition to the contributions from the neighboring transmitting devices. This conforms to the semantics of the proposed attacker behavior model that a device is likely to be compromised by an attacker if it has high vulnerability and is surrounded by devices that have high likelihood of being targeted. It can be argued that why not simply use the vulnerability of a device as a measure to estimate its likelihood of being compromised. The answer is in the fact that the position of a device in the NRG affects its likelihood of being reached by the attacker. If a highly vulnerable device is surrounded by devices with low vulnerabilities, the chances of the attacker reaching that device become less. Similarly, if a device with low vulnerability is surrounded by devices with high vulnerabilities then the likelihood of that device getting compromised increases. The rank and sampling probability computation module of TAS uses Algorithm \ref{algorithm1} to perform its operations. The algorithm produces two vectors $R = \{\rho_1, \rho_2, ..., \rho_n\}$ and $\Psi = \{\psi_1, \psi_2, ..., \psi_n\}$ corresponding to the vulnerability ranks and sampling probabilities of \emph{n} devices in the network. The values of $epochs$ and $thresh$ regulate the convergence conditions of the algorithm. For each device, first the vulnerability transference effect is computed and then it is combined with the vulnerability score to obtain the vulnerability rank value of the device. Finally, the vulnerability rank values are used to calculate the sampling probabilities by employing the softmax function. Algorithm \ref{algorithm1} is not computationally intense and it has been shown that PageRank for millions of web-pages can be computed in few hours on a medium size workstation \cite{google}. 

\begin{algorithm}[H]
\caption{Vulnerability Ranks and Sampling Probabilities Computation}
\label{algorithm1}
\begin{algorithmic}[1]
\renewcommand{\algorithmicrequire}{\textbf{Input:}}
\renewcommand{\algorithmicensure}{\textbf{Output:}}
\REQUIRE $NRG<D,E>, \delta, epochs, thresh$
\ENSURE $R = \{\rho_1, \rho_2, ..., \rho_n\}, \Psi = \{\psi_1, \psi_2, ..., \psi_n\}$
\STATE $R  \longleftarrow vulnerability \; scores \; of \; n \; devices \; in \; D$
\STATE $\Psi  \longleftarrow zero \; vector \; of \; length \; n$
\FOR{$i \in [1,epochs]$}
\STATE $R_{last} = R$
\FOR{$d_i \in D$}
\STATE /*vulnerability transference rank from Eq. \ref{rankeq}*/ 
\STATE $\tau_i = 1-\delta + \delta \sum_{j \in In(d_i)} \rho_j \frac{|Out(d_j, d_i)|}{|Out(d_j)|}$
\STATE $\rho_i = \rho_i + \tau_i$
\ENDFOR
\STATE /*calculate change in rank values using L1 norm*/ 
\STATE $err = sum(abs(R))-sum(abs(R_{last})) $
\IF{$err \leq thresh$}
\STATE break
\ENDIF
\ENDFOR
\FOR{$j \in [1,n]$}
\STATE /*softmax function to compute sampling probs.*/
\STATE $\psi_j = exp(\rho_j)/sum(exp(R))$ 
\ENDFOR
\end{algorithmic}
\end{algorithm}

\subsubsection{Realtime Probabilistic Sampling}
Once the vulnerability ranks and sampling probability computation module produces the vulnerability ranks $R$ and corresponding sampling probabilities $\Psi$ of all devices in the network, flows (or packets) are sampled in realtime before being sent to the zonal IDS. Let $X_{ij}$ be a flow from device $d_i$ to $d_j$, then the probability of flow $X_{ij}$ of being sampled is provided in Eq. \ref{sampeq}.

\begin{equation}
    P(X_{ij}) = \alpha \; \psi_i + \beta \; \psi_j 
    \label{sampeq}
\end{equation}

The sampling probabilities of devices $d_i$ and $d_j$ are given by $\psi_i,\psi_j \in \Psi$. The relative influence of source and destination device vulnerability ranks on the overall probability of sampling a flow is specified by $\alpha$ and $\beta$, where $\alpha,\beta \in [0,1]$ and $\alpha + \beta = 1$. A larger $\alpha$ value will sample more flows that have high vulnerability rank source devices, while a larger $\beta$ value will sample more flows with high vulnerability rank destination devices. In our experiments, the values of both $\alpha$ and $\beta$ are set to 0.5. It is possible to find optimal values of $\alpha$ and $\beta$ by employing parameter estimation techniques on historical intrusion detection data. The flows can be sampled in realtime with negligible overhead as the sampling probabilities of all devices are computed in the offline mode.

\subsection{Zonal Intrusion Detection System}
Zonal IDS receives sampled traffic from TAS, finds malicious activities in its surveillance zone and sends alerts to the ASM. IDSs can be either signature-based or anomaly-based, depending on the methodology they employ to detect intrusions. Signature-based IDSs maintain signatures or rules that are used to identify known cyber-attacks. These rules are generally created by the security analysts, but they can also be learned using supervised machine learning techniques. Signature-based IDS are widely used and perform well in identifying malicious traffic with considerable detail, but are unable to detect zero-day attacks. On the other hand, anomaly-based IDSs model the normal behavior of the network, and raise alerts when a traffic pattern is observed that deviates from the normal behavior. Anomaly-based IDSs can detect zero-day attacks, but they provide little information about the detected malicious activity. Anomaly-based IDSs are mostly developed using unsupervised machine learning techniques, and their use as a stand-alone IDS solution is limited. PRISM can incorporate any IDS provided that the alerts contain information regarding the vulnerability exploited, devices impacted and intrusion type. In its current implementation, PRISM uses the opensource signature-based Snort IDS \cite{snort}. However, PRISM can also include an anomaly-based IDS in synergy with a signature-based IDS in its surveillance zones. This can leverage PRISM in detecting anomalous traffic patterns leading to potential identification of zero-day attacks. The output of anomaly-based IDS can be used to calculate behavior-based network traffic metrics in the Security Analyzer system of PRISM introduced in Section 3.5 of the paper.     


\subsection{Alert Stream Manager}
ASM receives streams of alerts from the Zonal IDSs located in different surveillance zones, and forwards it to the Security Analyzer and Prediction Engine. The alerts are sent to the Security Analyzer as soon as they received. For security analysis functions, the order in which the alerts are received is not critical, but to update security information of the system promptly, it is essential to forward the alerts as soon as they are received. Between reception and forwarding of alerts to the Prediction Engine, ASM performs a set of important operations that ensures the alerts being forwarded to the Prediction Engine are in the same order as they are generated at their respective Zonal IDSs. The alerts can be reordered due to network latencies, unsynchronized clocks of Zonal IDSs and data transmission over a non-order-preserving channel. ASM addresses the alert reordering issue due to alerts being reported with different network latencies from different Zonal IDSs. To rectify this problem, ASM processes the unbounded and unordered streams of alerts by using a window-based stream management procedure. The essence of this procedure is the fact that we can estimate certain bounds on the network latencies. Bounds on the network latencies between Zonal IDSs and ASM can be computed from network characteristics or historical data that keeps track of the alert generation and reception time difference \cite{srivastava}.  


Consider $Z_1, Z_2, .., Z_n$ Zonal IDSs with alert reporting latencies upper-bounded by $L_1, L_2, .., L_n$, respectively. The actual maximum alert reporting latency is denoted by $\Delta_{max} = max(L_1, L_2, .., L_n)$, where the estimate of maximum alert reporting latency is represented by $\mu_{max}$. Each Zonal IDS generates a stream of alerts where an alert has a number of attributes, among which the alert generation timestamp $ts$ is crucial for ASM operations. The alerts are received by the alert buffer $Y = \{o_1, o_2, .., o_t,..\}$, where $o_1$ is first alert in the buffer and $o_t$ is the alert received at a certain time $t$. Alerts are processed and then sent to the Prediction Engine by a mechanism known as the Alert Order Rectification (AOR) procedure. The alert buffer occupancy is constantly being monitored and the number of alerts in the buffer at any instance is accounted by the variable \emph{buffer\_occupancy}. The interval between successive AOR procedure executions is also tracked and time elapsed since the last alert AOR procedure is tracked by the variable \emph{time\_elapsed}. The AOR procedure is triggered when \emph{buffer\_occupancy} surpasses a configurable parameter called \emph{alert sequence length} $\omega$ or \emph{time\_elapsed} exceeds another configurable parameter termed as \emph{timeout} $\eta$. The AOR procedure triggered when \emph{buffer\_occupancy} exceeds $\omega$ follows a three step protocol. First, a forwarding window $F$ is formed that includes all alerts in $Y$ and all alerts received after waiting for $\mu_{max}$ period of time. Second, the alerts in $F$ are sorted with respect to $ts$. Finally, the first $\omega$ number of alerts in $F$ are sent to the Prediction Engine and are removed from $Y$. The $\mu_{max}$ wait time ensures that all alerts that are supposed to be part of $F$ according to their alert generation timestamp, but were not received due to network latencies are included in $F$. It is possible that \emph{buffer\_occupancy} does not exceed $\omega$ for a long period of time. However, the alerts cannot remain in $Y$ indefinitely, and must be reported to the Prediction Engine for timely assessment of the attack. In that case, the AOR procedure is triggered when \emph{time\_elapsed} caps $\eta$, which also follows the three step protocol. The only difference is that all of the alerts in $Y$ are included in $F$, are sorted, and then sent to the Prediction Engine. As a result, a smaller sequence of alerts is sent to the Prediction Engine. The AOR procedure is articulated in Algorithm \ref{asmAlgorithm} and its time complexity is $O(n\:log(n))$, while $n$ here represents the number of alerts being sorted.

\begin{algorithm}[H]
\caption{Alert Order Rectification}
\label{asmAlgorithm}
\begin{algorithmic}[1]
\renewcommand{\algorithmicrequire}{\textbf{Input:}}
\REQUIRE $Y=\{o_1, o_2, ..\}, \mu_{max}, \omega, \eta$
\STATE \emph{buffer\_occupancy} $\longleftarrow no. \; of \; alerts \; in \; Y $
\STATE $time\_elapsed \longleftarrow time \; since \; last \; AOR \; proc.$
\WHILE{$Y \neq \emptyset$}
\IF {$(\emph{buffer\_occupancy} \geq \omega)$}
\STATE $F \longleftarrow  Y[o_1:o_{t+\mu_{max}}]$
\STATE $F.sort()$
\STATE $F = F[0:\omega]$
\STATE $Y = Y - F$
\STATE $sendToPredictionEngine(F)$
\STATE $F \leftarrow \emptyset$
\ENDIF
\IF {$(time\_elapsed == \eta)$}
\STATE $F \longleftarrow Y$
\STATE $F.sort()$
\STATE $Y = Y - F$
\STATE $sendToPredictionEngine(F)$
\STATE $F \leftarrow \emptyset$
\ENDIF
\STATE $update(\emph{buffer\_occupancy})$
\STATE $update(time\_elapsed)$
\ENDWHILE
\end{algorithmic}
\end{algorithm}

\par
\begin{figure} 
    \includegraphics[width=1\linewidth]{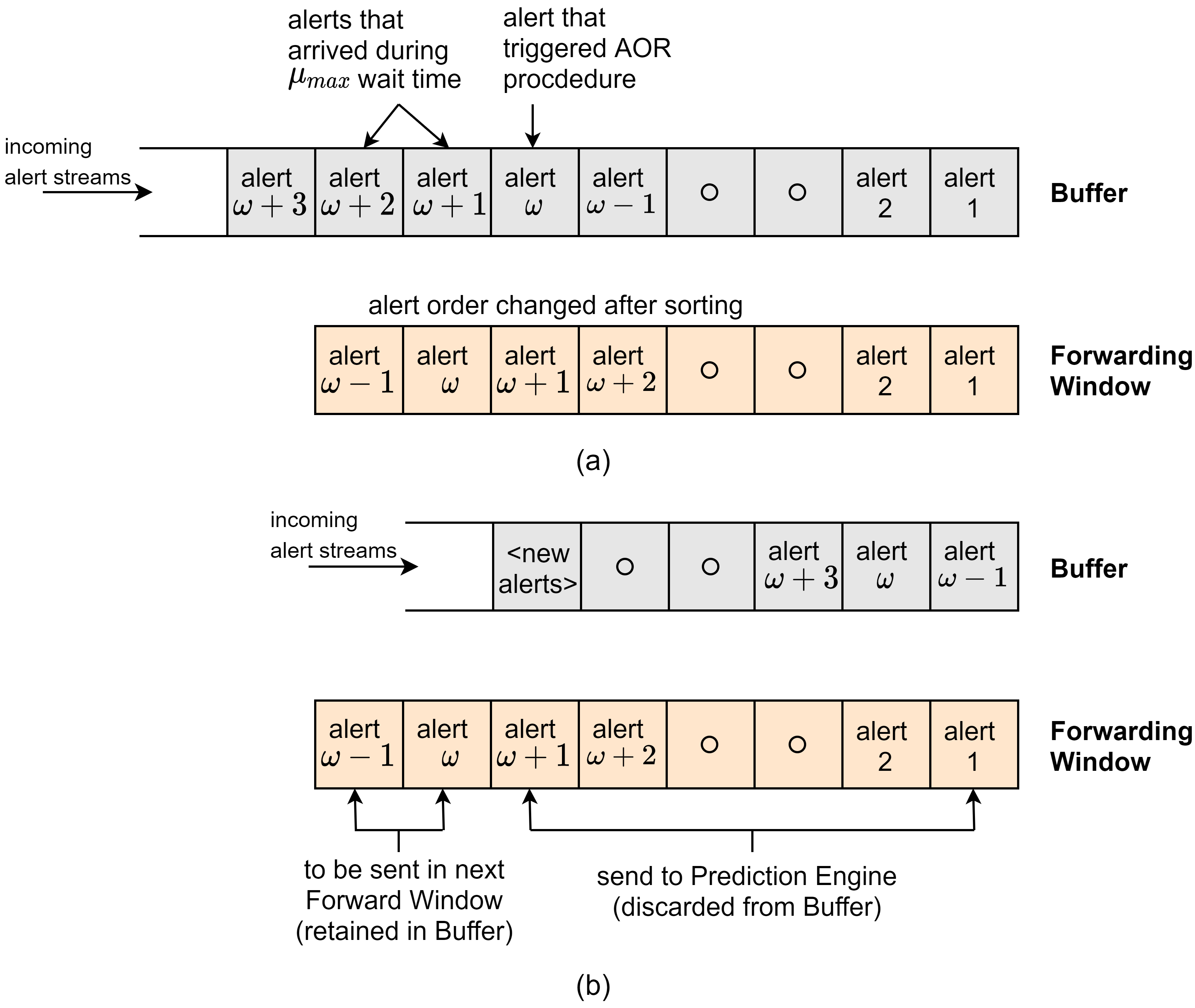}
  \caption{Evolution of buffer (Y) during the AOR procedure. (a) Creation of forwarding window (F) and sorting of alerts. (b) Sending alerts to Prediction Engine.}
  \label{aor} 
\end{figure}

 It is to be noted that for best results the value of $\mu_{max}$ needs to be as close as possible to $\Delta_{max}$. If we underestimate the value of $\mu_{max}$ then the rectification of alert reordering yields less precise results, and if the value of $\mu_{max}$ is overestimated then the original alert order can be restored by the ASM, but it will cause additional delay in sending alerts to the Prediction Engine. Also, the value set for $\omega$ controls the number of alerts to be processed by the Prediction Engine at a time. Generally, longer alert sequences allow more accurate attack prediction but at the cost of delay in attack prediction decision. The value of $\eta$ should be set considering the alert arrival rate. It should not be too small to trigger the timeout driven AOR procedure frequently and nor should be too large to introduce unnecessary delays in attack prediction. Fig \ref{aor} shows the state of $Y$ during the AOR procedure. The AOR procedure is triggered by buffer\_occupancy exceeding $\omega$, and correspondingly, $F$ is constructed by including all alerts that arrive until the $\mu_{max}$ time as shown in Fig \ref{aor}a. Subsequently, the first $\omega$ number of sorted alerts in $F$ are sent to the Prediction Engine and are discarded from $Y$ as illustrated in Fig. \ref{aor}b. 

\subsection{Prediction Engine}
Prediction engine performs the function of assessing the progression of an attack using the alerts forwarded by the ASM. PRISM has the flexibility to accommodate any time series data-based machine learning model, and in its current implementation the prediction engine is driven by HMM. Similar to TAS, prediction engine has offline and online modes of operation. In offline or training mode, prediction engine performs the operations of training data preprocessing and multi-stage attack learning. In online or prediction mode, attack stage prediction is carried out in realtime. The working of the internal modules of prediction engine is described as follows.

\begin{table}[!t]
\renewcommand{\arraystretch}{1.3}
\caption{Mapping of ATT\&CK Tactics to Prediction States}
\label{mappingTable}
\centering
\begin{tabular}{|l|l|}
\hline
\multicolumn{1}{|c|}{\textbf{ATT\&CK Tactics}} & \textbf{Prediction States}\\
\hline
i) Reconnaissance & \multirow{3}{*}{1) Initial access}\\
ii) Resource development & \\
iii) Initial access & \\
\hline
iv) Execution & 2) Execution\\
\hline
v) Persistence & \multirow{8}{*}{3) Foothold}\\
vi) Privilege escalation & \\
vii) Defense evasion & \\
viii) Credential access & \\
ix) Discovery & \\
x) Lateral movement &\\
xi) Collection & \\
xii) Command and control &\\
\hline
xiii) Exfiltration & \multirow{2}{*}{4) Impact}\\
xiv) Impact & \\
\hline
\end{tabular}
\end{table}

\subsubsection{Training Data Preprocessing}
The training of HMM utilizes IDS alert data that can be from IDS alerts of real attack instances, alerts from engineered attacks on a test network or synthetic alerts forged for training purposes. As the training data can be from multiple diverse sources, it is essential to formalize the preprocessing task. The training data preprocessing module accepts data in a CSV format then performs four step preprocessing operation. First, all rows with missing values are removed. Second, the timestamp format of alerts is checked and is reformatted if it is incompatible with the data parsing function. Third, duplicate rows are discarded. Fourth, the alert data is rearranged in the ascending order with respect to the timestamps. Finally, the refined IDS alert data is stored in a CSV file ready to be used by the multi-stage attack learning module of the prediction engine. 

\subsubsection{Multi-Stage Attack Learning}
The multi-stage attack learning module analyzes the preprocessed IDS alert data files to train the HMM. For each multi-stage attack type in the training data, a separate model is trained forming an HMM attack profile bank that is shared with the attack stage prediction module. PRISM uses the MITRE ATT\&CK \cite{mitre} adversary tactics and techniques knowledge base to contrive the state space of its prediction models that correspond to different attack progression stages. ATT\&CK is a real-world observations-based threat information framework that describes the actions of adversaries in executing complex multi-stage attacks. These actions are represented by the techniques in ATT\&CK framework that are organized into fourteen tactics. ATT\&CK provides a detailed technical description, real-world usage examples with associated actors, mitigation and detection information for each technique. Tactics are the tactical objectives of the adversary that are achieved by employing the relevant techniques and tactics are arranged in the way attacker progresses through a multi-stage attack. The tactics in ATT\&CK framework naturally become suitable candidates for the states in the multi-stage attack prediction models. However, having fourteen states not only adds to computational complexity, but having all fourteen tactics involved in a single attack is highly unlikely. This necessitates the mapping of fourteen ATT\&CK tactics into a smaller and more generic state space. The prediction engine of PRISM maps the fourteen tactics into four states: initial access, execution, foothold, and impact. The mapping between ATT\&CK tactics and the prediction states is shown in Table \ref{mappingTable}. Note that the terms attack stage and prediction state are used interchangeably in the paper. As discussed earlier, a supervised learning approach is used to train the HMM. The preprocessed IDS alert training data does not have labels indicating the classification of alert sequence into ATT\&CK tactics and corresponding prediction states. Most of the mainstream threat detection frameworks have the feature to report alerts in compliance to the ATT\&CK framework \cite{splunk, extrahop, logrhythm}. Since PRISM relies on the opensource snort IDS, a mechanism is required to associate snort alerts to the ATT\&CK tactics. For this purpose, we have utilized the concepts of information retrieval to match the key words from snort alert messages to the technical description of the ATT\&CK techniques which is discussed in detail in Section 4 of the paper. Using the labelled IDS alert training data, $A$ and $B$ matrices are learned using Eqs. \ref{AmatrixLearn} and \ref{BmatrixLearn}.



\begin{figure}[!t]
\centering
\includegraphics[width=3.4in]{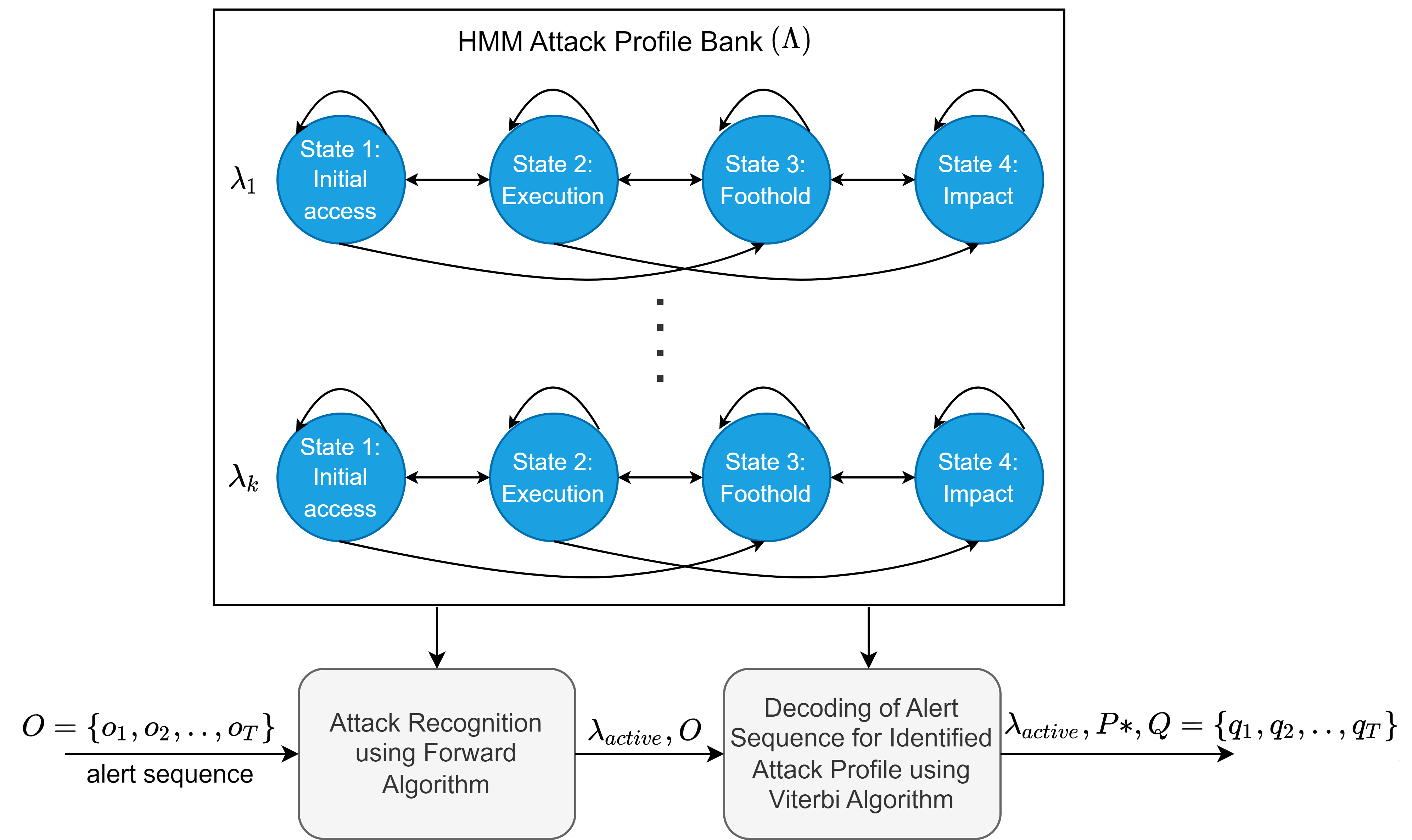}
\caption{The process of attack recognition and decoding.}
\label{fig:prediction_process}
\end{figure}

\begin{equation}
\begin{split}
    & a_{ij} =  \frac{\Gamma_{ij}}{\sum_{r=1}^{N}\Gamma_{ir}}, \;\;\;\;\;\;\; i,j \in [1, N] \\
    & s.t. \;\;\;\;\; a_{ij} \; = \; 0, \;\;\;\;\;\;\;\;\;  j > i+2 \\
    & \;\;\;\;\;\;\;\;\;\;\;\;\;\;\;\;\;\;\;\;\;\;\;\;\;\;\;\;\;\;\;\;\; j < i-1  
\end{split}
\label{AmatrixLearn}
\end{equation}



\begin{equation}
\begin{split}
    b_i(l) = \frac{\Upsilon_i(l)}{\sum_{r=1}^{M}\Upsilon_i(r)}, \;\;\;\;\;\;\; & i \in [1, N] \\
    & l \in [1, M]
\end{split}
\label{BmatrixLearn}
\end{equation}

\par Where, $\Gamma_{ij}$ is the number of transitions from state $s_i$ to $s_j$ and $\Upsilon_i(l)$ is the number of times observation $v_l$ appears in state $s_i$. The learned HMM can be ergodic or of any shape based on the state transitions present in the training data. However, there are certain HMM types that model some applications more accurately, e.g., the left-right model is considered to have better performance in speech recognition \cite{HMMRabiner}. Based on our experimentation, we have devised a semi-ergodic HMM for PRISM. The transition probabilities being learned in Eq. \ref{AmatrixLearn} are constrained by the two conditions that no transitions of more than two states are permitted and transitions from higher states to lower states are only possible among the adjacent states. These constraints improve the accuracy in detecting multi-stage attack progression. Algorithm \ref{algorithm2} explains the multi-stage attack learning process where it receives the alert training data and produces the HMM attack profile bank $\Lambda$. It is to be noted that the models are designed to always begin from the first state, that is why all models in $\Lambda$ are initialized with $\pi_{i} = [1\;0\;0\;0]$.

\begin{algorithm}[H]
\caption{Multi-stage Attack Learning}
\label{algorithm2}
\begin{algorithmic}[1]
\renewcommand{\algorithmicrequire}{\textbf{Input:}}
\renewcommand{\algorithmicensure}{\textbf{Output:}}
\REQUIRE $Training\_Data = \{attack_1, attack_2, .., attack_k\}$
\ENSURE $\Lambda = \{\lambda_1, \lambda_2, .., \lambda_k \}$
\STATE $\Lambda \longleftarrow vector \; of \; k \; \lambda \; objects \; (init. \: with \: \pi_{i})$
\FOR{$x \in [1,k]$}
\STATE /*Learning Transition Probabilities*/
\FOR {$i \in [1,N]$}
\FOR {$j \in [1,N]$}
\STATE $\lambda_x.A[i][j] = \frac{\Gamma_{ij}}{\sum_{r=1}^{N}\Gamma_{ir}}$ /* from Eq. \ref{AmatrixLearn}*/
\IF {($j > i+2 \; || \; j < i-1$)}
\STATE $\lambda_x.A[i][j] = 0$
\ENDIF
\ENDFOR
\ENDFOR
\STATE /*Learning Emission Probabilities*/
\FOR {$i \in [1,N]$}
\FOR {$l \in [1,M]$}
\STATE $\lambda_x.B[i][j] = \frac{\Upsilon_i(l)}{\sum_{r=1}^{M}\Upsilon_i(r)}$ /* from Eq. \ref{BmatrixLearn}*/
\ENDFOR
\ENDFOR
\ENDFOR
\end{algorithmic}
\end{algorithm}

\subsubsection{Attack Recognition and Stage Prediction}
The attack recognition and stage prediction module predicts the type of the multi-stage attack and its current stage for every alert sequence forwarded by the ASM in realtime. The attack stage prediction process has two parts, attack recognition and decoding. In attack recognition, the attack profile corresponding to the observed alert sequence is determined from the HMM attack profile bank. To find out the most likely attack profile, the likelihood of the alert observation sequence given the model, $P(O|\lambda)$, is computed for each HMM attack profile using the forward algorithm. The HMM attack profile with the highest likelihood score is selected and is used to track the progress of the attack. The mechanics of forward algorithm are discussed in Appendix A of the paper. After attack recognition, the process of decoding takes place that involves finding out the optimal state sequence corresponding to the forwarded alert sequence. To achieve this, the variable $\chi_t(i)$ is introduced that represents the probability of being in state $s_i$ at time $t$ as expressed in Eq. \ref{previterbi1}. The most likely state $q_t$ at time $t$ can be solved independently using Eq. \ref{previterbi2}. Viterbi algorithm finds $q_t$ for a given observation sequence efficiently using a dynamic programming approach as explained in Appendix A of the paper.

\begin{equation}
    \chi_t(i) = P(q_t = s_i | O,\lambda)
\label{previterbi1}
\end{equation}

\begin{equation}
    q_t(i) = \underset{i = 1, ..,N}{\argmax} \chi_t(i), \;\;\; t \in [1, T]
\label{previterbi2}
\end{equation}

Algorithm \ref{algorithm3} describes the functioning of attack recognition and stage prediction module. The time complexity of Algorithm \ref{algorithm3} is $O(N^2T)$ which is the time complexity of both forward and Viterbi algorithms. Fig. \ref{fig:prediction_process} depicts the use HMM attack profile bank in the attack recognition and decoding tasks.

\begin{algorithm}[H]
\caption{Attack Recognition and Stage Prediction}
\label{algorithm3}
\begin{algorithmic}[1]
\renewcommand{\algorithmicrequire}{\textbf{Input:}}
\renewcommand{\algorithmicensure}{\textbf{Output:}}
\REQUIRE $O=\{o_1, o_2, .., o_T\},\; \Lambda = \{\lambda_1, \lambda_2, .., \lambda_k\}$
\ENSURE $\lambda_{active},\; P*,\; Q = \{q_1, q_2, .., q_T\}$
\STATE /* Attack Recognition by Forward Algorithm */ 
\STATE max\_likelihood = 0
\FOR{$\lambda \in \Lambda$}
\STATE likelihood = $P(O|\lambda)$ /* from Eq. A.3 */
\IF {($likelihood > max\_likelihood$)}
\STATE max\_likelihood = likelihood
\STATE $\lambda_{active} = \lambda$ 
\ENDIF
\ENDFOR
\STATE /* Decoding using Viterbi Algorithm */
\STATE $\nu_t(i) = \underset{q_1,.., q_{t-1}}{\max} P(q_1, .., q_{t-1},o_1, .., o_t, q_t = i | \lambda_{active})$
\STATE $P*$ = $\underset{i = 1, ..,N}{\max} \nu_T(i)$ /* from Eq. A.10 */
\STATE $Q'$ = $\underset{i = 1, ..,N}{\argmax} \; \nu_T(i)$ /* from Eq. A.11 */
\FOR{$t = T-1:1$}
\STATE $Q' =  Q' \cup \xi_{t+1}(q_{t+1}^*)$
\ENDFOR
\STATE $Q = Q'.reverse()$
\end{algorithmic}
\end{algorithm}

\subsection{Security Analyzer}
Security Analyzer realizes the cyber situational awareness capability of PRISM by the computation of important security metrics and their visualization in realtime for effective attack response. The input to the Security Analyzer is prediction output of the Prediction Engine and unmediated alerts from the surveillance zones. By unmediated it is meant that the alerts are not subjected to AOR procedure in the ASM. The alert and prediction output data is processed in realtime to extract relevant information for security metrics computation. The information of interest in the alert data is the alert id, alert generation timestamp and id of the device for which the alert has been generated. From the prediction output, information regarding attack type, attack stage and the probability of being at different attack stages corresponding to an alert is obtained. The metric computation process is alert-driven (event-driven), and the metrics are computed with each reported alert at time $t$. To get a holistic picture of the damage spread, widely used system availability metric is utilized. Additionally, we propose two metrics: threat perceptivity and system degradability that are designed to enhance the cyber situational awareness capabilities specifically for the case of availability attacks. Formally, the three metrics are defined as follows.

\par \textbf{System availability: }The percentage of devices working at their routine operational capacity with respect to the total number of devices in the system at a certain time $t$. It is expressed in Eq. \ref{sysAvail}. 

\begin{equation}
    \emph{system availability(t)} = \frac{\emph{no. of available devices at t}}{\emph{total no. of devices}} \:*\: 100
\label{sysAvail}
\end{equation}

\par \textbf{Threat perceptivity: }The measure to quantify attack progression and associated risk to the system. Mathematically, it is expressed in Eq. \ref{tPercept}, where $\chi_t(i)$ is the probability of being at state $s_i$ for an alert received at time $t$, $\varepsilon_i$ is the numerical value corresponding to the risk associated with state $s_i$, with higher values to be set for advanced stages. The value of thereat perceptivity is between 0 and 1 and higher value means the ongoing attack is in advanced stages with the system facing elevated risk.

\begin{equation}
    \emph{threat perceptivity(t)} = \frac{\sum_{i=1}^N \chi_t(i) \varepsilon_i}{max(\varepsilon_i)}
\label{tPercept}
\end{equation}

\par \textbf{System degradability: }The impact of attack risk on the system operability quantified through the fraction of compromised devices in the system. Eq. \ref{sysDeg}a specifies the fraction of compromised devices at time $t$, and system degradability at time $t$ is manifested in Eq. \ref{sysDeg}b. Like threat preceptivity, the value of system degradability is also between 0 and 1, with higher values implying increased attack progression intensity.

\begin{subequations}
\begin{align}
& \theta(t) \; = \; \frac{\emph{no. of compromised devices at t}}{\emph{total no. of devices}} \\
& \emph{system degradability(t)} = \theta(t) \; \emph{threat perceptivity(t)} 
\end{align}
\label{sysDeg}
\end{subequations}

System availability, threat perceptivity and system degradability are computed for each reported alert and the Security Analyzer plots the metrics in realtime for the convenience of the security operations team to gain valuable insights about the ongoing attack. Visualization of the metrics helps to assess the situation better and consequently a more informed response can be deployed. Additionally, all of the metric computations are constantly being stored in a report file that can be used by an automated response system. The visualization functionality of the Security Analyzer is discussed further in Section 5 of the paper.    

%
%

\section{Experimental Setup}
\begin{figure}[!t]
\centering
\includegraphics[width=3.5 in]{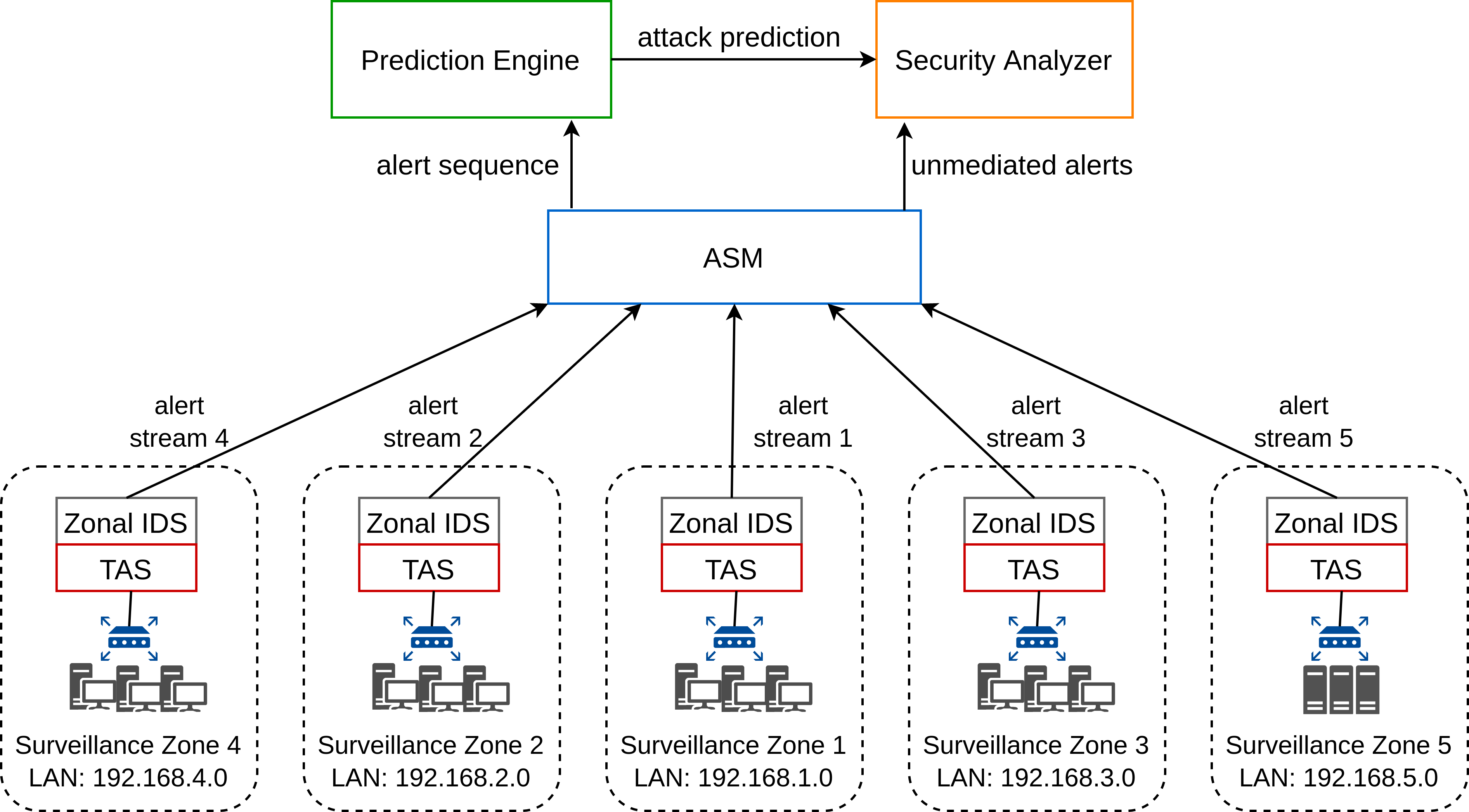}
\caption{PRISM deployment on the ISCX-2012 network.}
\label{arch-cic}
\end{figure}

To evaluate the performance of PRISM, extensive experimentation has been conducted on the ISCX-2012 and CIC-2017 datasets. The ISCX-2012 dataset incorporates several multi-stage attacks launched on a widespread enterprise network while the CIC-2017 dataset is more considerate about accurate characterization of normal and attack traffic behavior profiles being appraised on a small-scale network. Therefore, the ISCX-2012 dataset has been utilized to investigate the utility of distributed security monitoring functionality of PRISM in addition to evaluating multi-stage attack prediction performance and the effectiveness of PRISM in realizing its cyber situational awareness capability. To examine the robustness of multi-stage attack prediction mechanism of PRISM in its ability to detect a multi-stage attack and correctly predict its progression, experimentation on the CIC-2017 dataset is conducted. All constituent systems of PRISM are implemented in Python 3.7 and experiments are conducted on a workstation with 8 cores and 16 GB of memory. The specifications of different attacks used in our experimentation are provided in Table \ref{attackSpec}. The infiltration attack scenario depicts how an attacker compromises a web server and associated database by gaining an initial foothold and then moving laterally in the network. In the HTTP DoS attack, the attacker gains an initial foothold in the network and after compromising several devices, launches a Slowloris HTTP DoS attack on the web server. The DDoS botnet attack scenario illustrates how the attacker deploys Ares botnet tool on vulnerable devices and launches DDoS attack on the web server using Low Orbit Ion Canon (LOIC) DDoS attack tool. Fig. \ref{arch-cic} shows the conceptual deployment of PRISM on ISCX-2012 network architecture by realizing each Local Area Network (LAN) as a surveillance zone. As the network of CIC-2017 dataset contains only 10 devices connected through a single LAN, therefore only one surveillance zone is established for the experiments related to the CIC-2017 dataset. The details of how different modules of the constituent systems of PRISM implement the distributed security monitoring and integrated security analysis functionality using the datasets' resources are discussed as follows.

\begin{table*}[!t]
\renewcommand{\arraystretch}{1.3}
\caption{Attack Specifications}
\label{attackSpec}
\centering
\begin{tabular}{|l|c|c|l|}
\hline
\multicolumn{1}{|c|}{\textbf{Attack Name}} & \multicolumn{1}{c|}{\textbf{Dataset}} & \multicolumn{1}{c|}{\textbf{No. of Packets}} & \multicolumn{1}{c|}{\textbf{Attack Description}}\\
\hline
\multirow{6}{*}{Infiltration Attack} & \multirow{6}{*}{ISCX-2012} & \multirow{6}{*}{5,763,149} & 1) Gather information about the target including IP ranges and user email accounts  \\
& & & 2) Send phishing emails to target user accounts with malicious pdf file attached  \\
& & & 3) User opens malicious file causing arbitrary code execution \\
& & & 4) From the compromised device more devices in the network are targeted  \\
& & & 5) Network scanning is performed from infected devices to identify web servers \\
& & & 6) Web server is targeted using injection techniques causing its complete takeover \\
\hline
\multirow{6}{*}{HTTP DoS Attack} & \multirow{6}{*}{ISCX-2012} & \multirow{6}{*}{9,648,653} & 1) Reconnaissance using nslookup and sending phishing emails with malicious file\\
& & & 2) Victim device is infected through arbitrary code execution from malicious pdf file\\
& & & 3) More vulnerable devices are discovered through network scanning \\
& & & 4) Devices are compromised by exploiting vulnerable SMB authentication protocol  \\
& & & 5) Python-based DoS attack tool Slowloris is deployed on the compromised devices \\
& & & 6) Slowloris HTTP DoS attack is launched on the web server from 4 infected devices  \\
\hline
\multirow{4}{*}{DDoS Botnet Attack} & \multirow{4}{*}{CIC-2017} & \multirow{4}{*}{9,997,874} & 1) Portscan of the target network is carried out\\
& & & 2) Ares botnet tool on 5 vulnerable devices is deployed \\
& & & 3) More vulnerable devices are discovered through network scanning \\
& & & 4) LOIC-based DDoS attack is launched on the web server\\
\hline
\end{tabular}
\end{table*}

\subsection{Distributed Security Monitoring}
TAS and Zonal IDS implement the distributed security monitoring functionality of PRISM. As mentioned in Section 3, TAS performs the operations of network reachability graph generation, vulnerability assessment, vulnerability ranks and sampling probabilities computation, and realtime probabilistic sampling. The network reachability graph generation is carried out by determining the communication between all network devices on different port numbers using the normal network activity data in the dataset. For vulnerability assessment, the vulnerability score of each device is estimated as we cannot deploy vulnerability scanners on the dataset networks nor vulnerability information of the network devices is provided in the datasets. To estimate the vulnerability scores of devices, we use the CVSS base score computation methodology by treating network devices as vulnerabilities to be exploited. The CVSS base score is determined using five exploitability metrics: attack vector, attack complexity, privileges required, user interaction and scope, along with three impact metrics: confidentiality, integrity and availability. The attack vector is determined using network reachability graph with the assumption that all devices that can communicate to a device can also compromise it. Therefore, the part of the network from where a device could potentially be targeted can be determined. Since we are modeling multi-stage attacks, the attack complexity is set to \emph{high} for each device. Privileges required is set as \emph{low} for regular hosts and \emph{high} for servers. User interaction is set as \emph{required} and scope is set as \emph{changed} for each device. For all three impact metrics, the value set to \emph{low} for regular hosts and \emph{high} for servers. Table \ref{cvssTable} summarizes the values adjusted for CVSS base score metrics. Once the network reachability graph is generated and the vulnerability assessment process is complete, the vulnerability ranks and sampling probabilities of all devices in the network are calculated using Algorithm \ref{algorithm1}. The network traffic incorporated in pcap files is then distributed between the surveillance zones using the IP addresses of the devices. In each surveillance zone, the packets in the zonal pcap file are sampled using Eq. \ref{sampeq}. The sampled zonal pcap file is then forwarded to Snort IDS that generates an alert file using Snort v3.0 community rules. To engineer the distributed alert reporting process, the alert file in each surveillance zone is appended with a transmission delay value between $\Delta_{min}$ and $\Delta_{max}$, which are configurable parameters in our experiments, and for each surveillance zone, the transmission delay values are randomly chosen within the range of these parameters. The alerts from each surveillance zone are then delivered to the input buffer of the ASM according to the alert generation timestamp added with the transmission delay.

\begin{table}[!t]
\renewcommand{\arraystretch}{1.3}
\caption{CVSS Metrics Values for Vulnerability Assessment}
\label{cvssTable}
\centering
\begin{tabular}{|c|c|c|}
\hline
\textbf{Metric Name} & \textbf{Metric Value}\\
\hline
Attack Vector & Determined by NRG \\
\hline
Attack Complexity & High \\
\hline
Privileges Required & Regular hosts: low, Servers: high \\
\hline
User Interaction & Required \\
\hline
Scope & Changed \\
\hline
Confidentiality & Regular hosts: low, Servers: high \\
\hline
Integrity & Regular hosts: low, Servers: high \\
\hline
Availability & Regular hosts: low, Servers: high \\
\hline
\end{tabular}
\end{table}

\subsection{Integrated Security Analysis}
The integrated security analysis functionality of PRISM is implemented by ASM, Prediction Engine and Security Analyzer. Initially, the multi-stage attack learning operation of the Prediction Engine is performed to construct the HMM attack profile bank. The pcap files for different attack types in the datasets are used to generate the alert files using Snort IDS. The four-step training data preprocessing is carried out and then the alert labelling corresponding to ATT\&CK tactics is accomplished. To label an alert, keywords are extracted from the alert information and are matched to the technical description of ATT\&CK techniques using the information retrieval method of Term Frequency - Inverse Document Frequency (TF-IDF). The TF-IDF scores of each keyword are calculated for all ATT\&CK techniques and the technique that has the highest average TF-IDF score of the alert message keywords is identified. The ATT\&CK tactic that pairs to the identified ATT\&CK technique is selected, and prediction state corresponding to the selected tactic, as described in Table \ref{mappingTable}, is chosen as the label of the alert. Some ATT\&CK techniques are part of multiple tactics, therefore, labelling of alerts requires establishment of context. The context is established by using the state corresponding to the last labelled alert. For multiple candidate tactics picked up for an alert, those tactics are selected that maps to a higher prediction state, otherwise the alert is labelled with the state of the last labeled alert. The semantics behind this rule is that as the attacker is executing a multi-stage attack, it is counter intuitive for the attacker to move in the backwards direction. If there are several alerts pointing to the previous prediction states then that will be considered as a new multi-stage attack. Such attack scenarios are investigated in \cite{shawly}. In rare cases, the average TF-IDF score of alert message keywords is not significantly different for different ATT\&CK techniques due to limitation of expressiveness in Snort alert information. To handle such cases we label the alert with the label of the previous alert, i.e., we make the assumption that this alert is not causing any attack stage transition. Once the alert files are labelled, Algorithm \ref{algorithm2} trains the HMM attack profiles with 30\% of the alert data of each attack type used in the experimentation.

\par Alerts from different surveillance zones are gathered in the input buffer of the ASM, from where they are sent to the Prediction Engine and Security Analyzer. The alerts being sent to the Prediction Engine are first sequenced and then forwarded using the AOR procedure articulated in Algorithm \ref{asmAlgorithm}. The value of the parameter $\mu_{max}$ is configured to be less than the value set for $\Delta_{max}$ in the experiments to test the robustness of alert stream management process. The parameter $\omega$ is selected according to the maximum length of alert sequence intended to be processed by the Prediction Engine. The value of the $\eta$ parameter is fixed to be 500 ms in all experiments. Prediction Engine upon reception of an alert sequence first finds out which attack is active and then predicts the stage of the attack using Algorithm \ref{algorithm3}. Security Analyzer receives unmediated alerts from the ASM and attack prediction information from the Prediction Engine to compute and visualize three metrics: system availability, threat perceptivity and system degradability for each incoming alert.

\section{Performance Evaluation}
We have conducted several experiments to evaluate the performance of PRISM in terms of processing efficiency and prediction efficacy. The experiments are designed to demonstrate the individual performance of the constituent systems of PRISM and their influence on the overall system operability.

\subsection{Effect of Threat-Aware Sampling and Distributed Traffic Processing}
\begin{figure} 
    \centering
  \subfloat[\label{exp1a}]{%
       \includegraphics[width=0.5\linewidth]{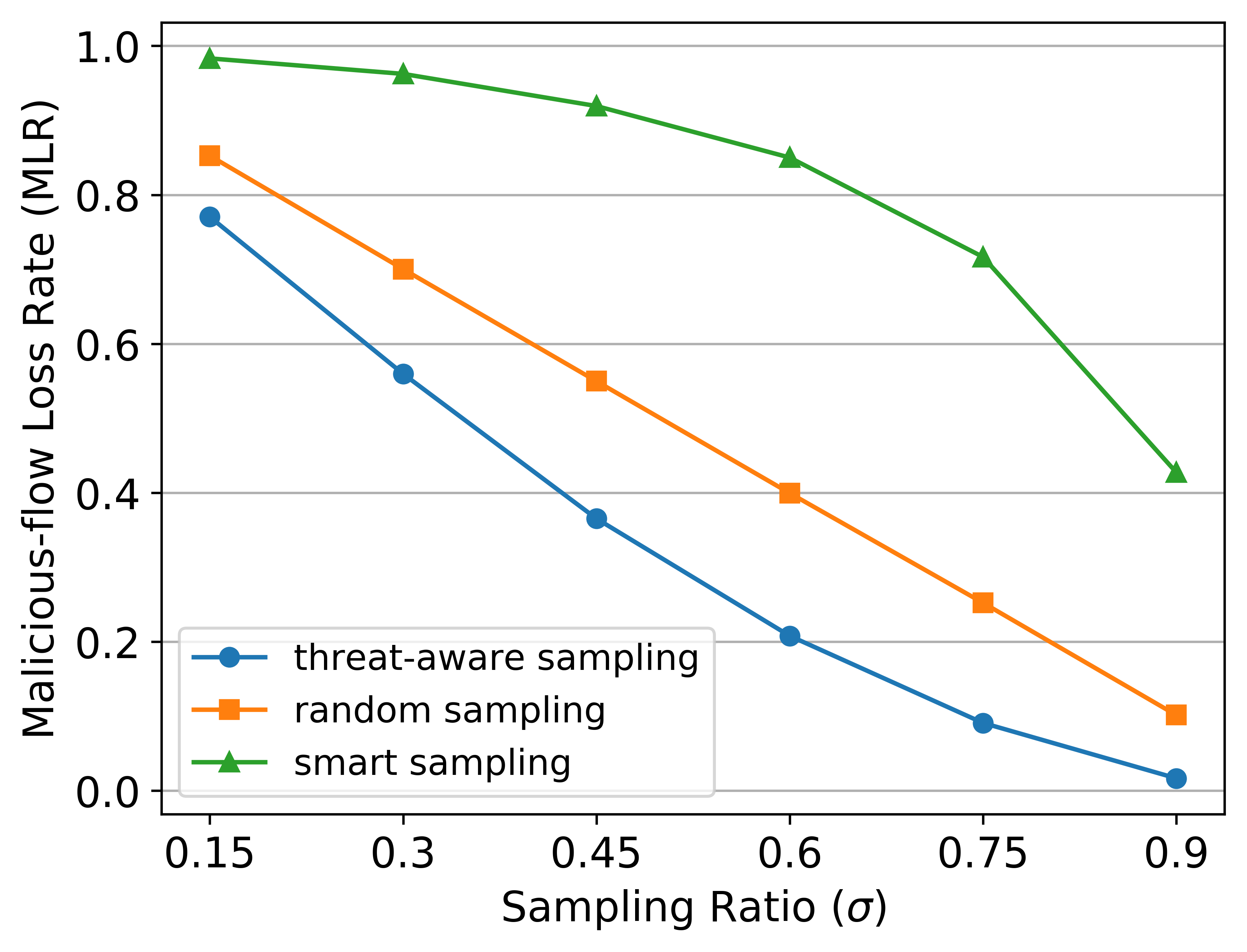}}
    \hfill
  \subfloat[\label{exp1b}]{%
        \includegraphics[width=0.5\linewidth]{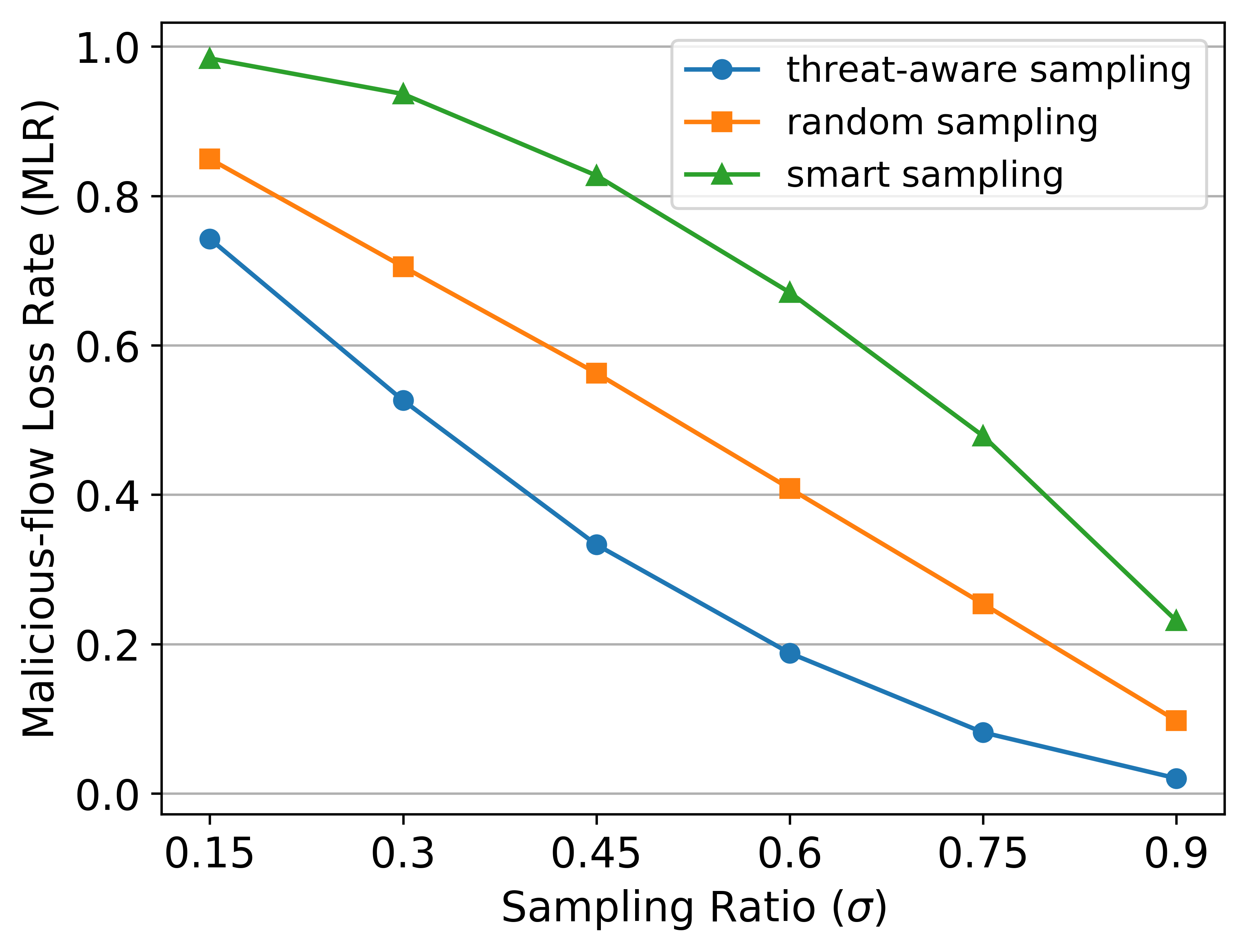}}
  \caption{Threat-aware sampling performance in comparison to common sampling schemes for (a) infiltration attack, and (b) HTTP DoS attack.}
  \label{exp1} 
\end{figure}

The most computation intensive phase in the intrusion detection exercise is the processing of network traffic. The vast magnitude of traffic generated by various devices in the enterprise networks makes it even more challenging, especially when the requirement is to detect malicious traffic in realtime. To cope up with this challenge, a simple yet expensive solution is commonly employed, that is, adding more computation power. PRISM on the other hand provides an effective solution that leverages its threat-aware sampling and distributed traffic monitoring structure to process traffic in realtime with limited computation resources. In the first experiment, the performance of threat-aware sampling is compared to common network traffic sampling schemes of smart sampling and random sampling. In smart sampling, the flows are sampled according to their size with large flows having more probability of being selected \cite{duffield}. The comparison between the sampling schemes is made using a measure that determines how much information is lost in the sampling process. In intrusion detection applications, a malicious flow is the information that is desired to be retained where a normal flow is considered as noise. We introduce the metric Malicious-flow Loss Rate (MLR) that determines the amount of malicious flows not retained after sampling as expressed in Eq. \ref{mlr}. Sampling ratio $\sigma$ is the control parameter in the threat-aware sampling process, which is the number of flows selected over the total number of flows as manifested in Eq. \ref{sigma}. The MLR of threat-aware sampling, random sampling and smart sampling corresponding to different values of $\sigma$ for infiltration and HTTP DoS attacks is shown in Fig. \ref{exp1}. It can be seen that the MLR of threat-aware sampling is much lower than that of random sampling and smart sampling with the widest difference observed for $\sigma=0.45$ and $\sigma=0.6$ for both attack types. The MLR values of all sampling schemes for both attack types are close to each other when $\sigma=0.15$ and $\sigma=0.90$. The cause of this behavior relies on the fact that when the sampling ratio is low, only few flows have to be selected out of the total, that forces the sampling schemes including the threat-aware sampling to lose malicious flows. On the other hand, when the sampling ratio is high, only few of the flows are needed to be dropped that leaves all sampling schemes to retain most of the malicious flows.  

\begin{equation}
    MLR = \frac{\emph{number of malicious flows not selected}}{\emph{total number of malicious flows}}
\label{mlr}
\end{equation}

\begin{equation}
    \sigma = \frac{\emph{number of sampled flows}}{\emph{total number of flows}}
\label{sigma}
\end{equation}

\begin{figure} 
    \centering
  \subfloat[\label{exp2a}]{%
       \includegraphics[width=0.5\linewidth]{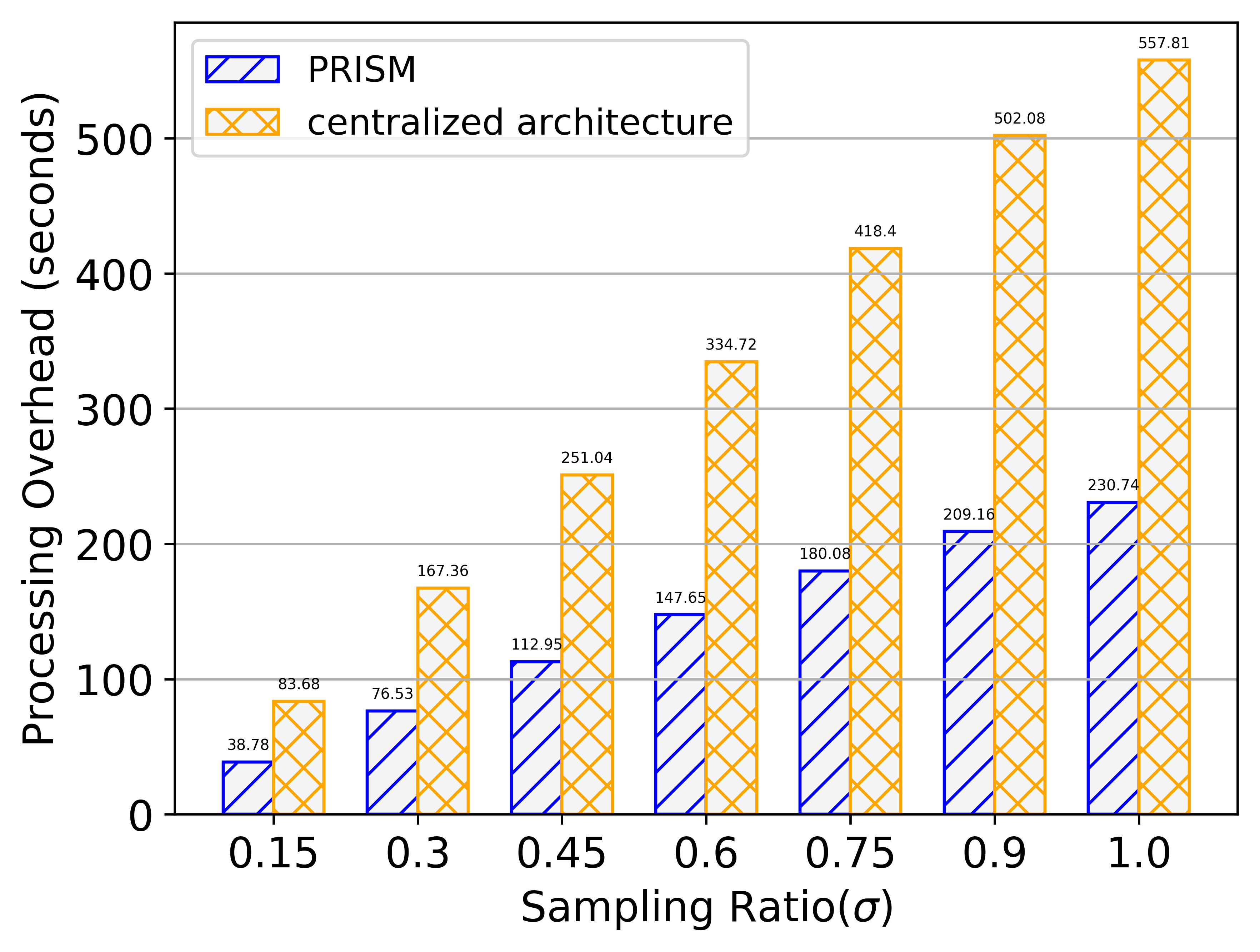}}
    \hfill
  \subfloat[\label{exp2b}]{%
        \includegraphics[width=0.5\linewidth]{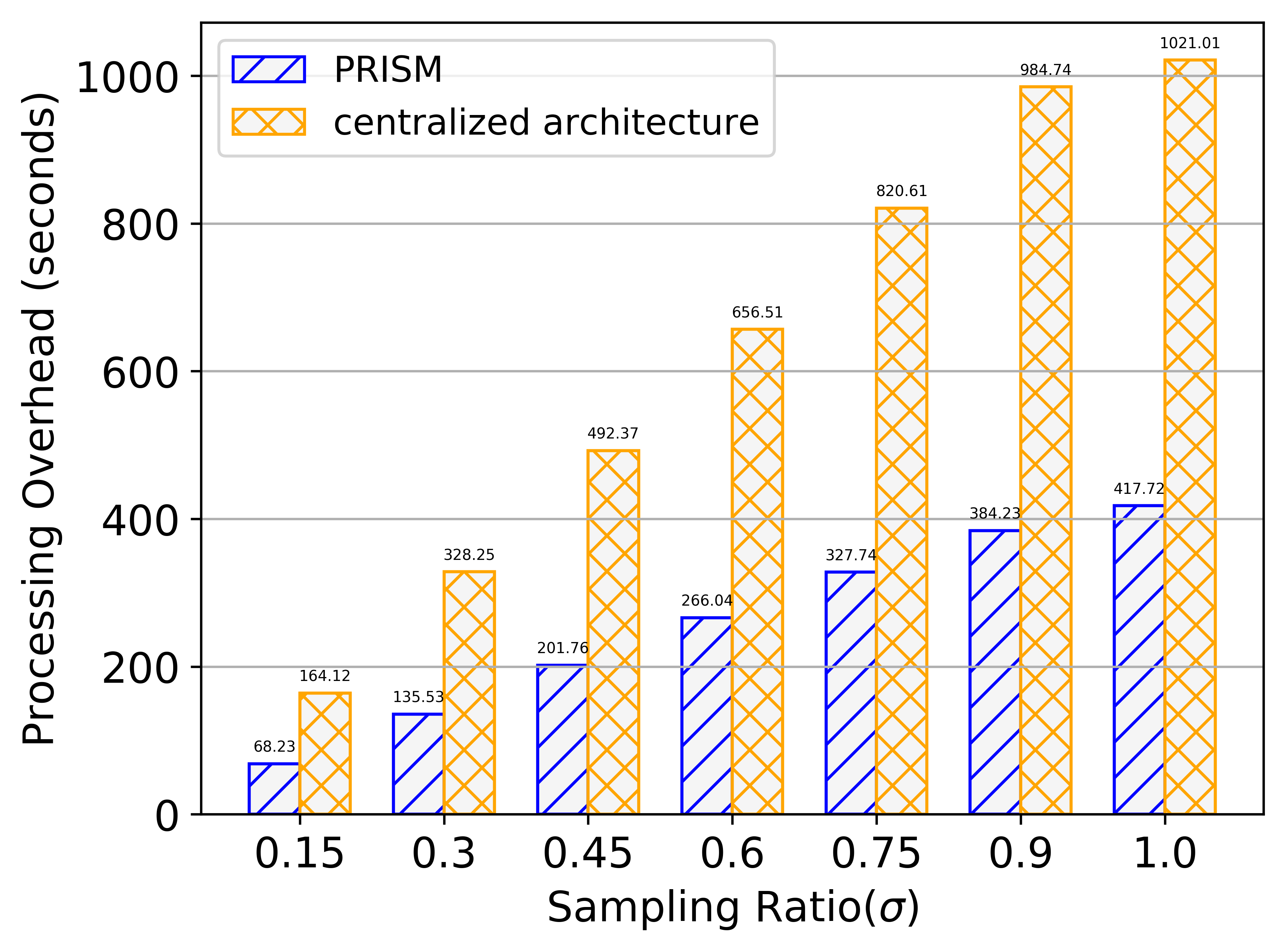}}
  \caption{Processing overhead of centralized intrusion detection and distributed security monitoring of PRISM corresponding to different sampling ratios for (a) infiltration attack, and (b) HTTP DoS attack.}
  \label{exp2} 
\end{figure}

In the second experiment, we have investigated the effect of threat-aware sampling and distributed architecture on the traffic processing capabilities of Snort. Fig. \ref{exp2} shows the processing overhead of Snort for around 5.8 million packets of infiltration attack data and 9.6 million packets of HTTP DoS attack data. As expected, the performance of PRISM's distributed architecture is significantly better than a centralized intrusion detection architecture for different sampling ratios. It has been determined in the experiments presented in Section 5.2 of the paper that the prediction capabilities of PRISM are acceptable with even $\sigma=0.3$. Comparing the processing overhead of PRISM's distributed architecture setting with $\sigma=0.3$ to the processing overhead of non-distributed and no sampling system ($\sigma=1$) as an equivalent to a standard IDS (like Snort) shows that PRISM makes the intrusion detection process 7.3x faster in the case of infiltration attack and 7.5x faster in HTTP DoS attack scenario.


\begin{table}[!t]
\renewcommand{\arraystretch}{1.3}
\caption{Distributed Alert Reporting Parameters}
\label{distAlert}
\centering
\begin{tabular}{|c|c|c|}
\hline
\textbf{Delay Configuration} & $\boldsymbol{\Delta_{min}-\Delta_{max}}$ \textbf{(ms)} & $\boldsymbol{\mu_{max}}$ \textbf{(ms)}\\
\hline
Low & 50 - 100 & 70 \\
\hline
Medium & 100 - 250 & 175 \\
\hline
High & 250 - 400 & 280 \\
\hline
\end{tabular}
\end{table}

\subsection{Attack Prediction Performance and Utility of Alert Stream Management}
\begin{figure} 
    \centering
  \subfloat[\label{2exp1a}]{%
       \includegraphics[width=0.5\linewidth]{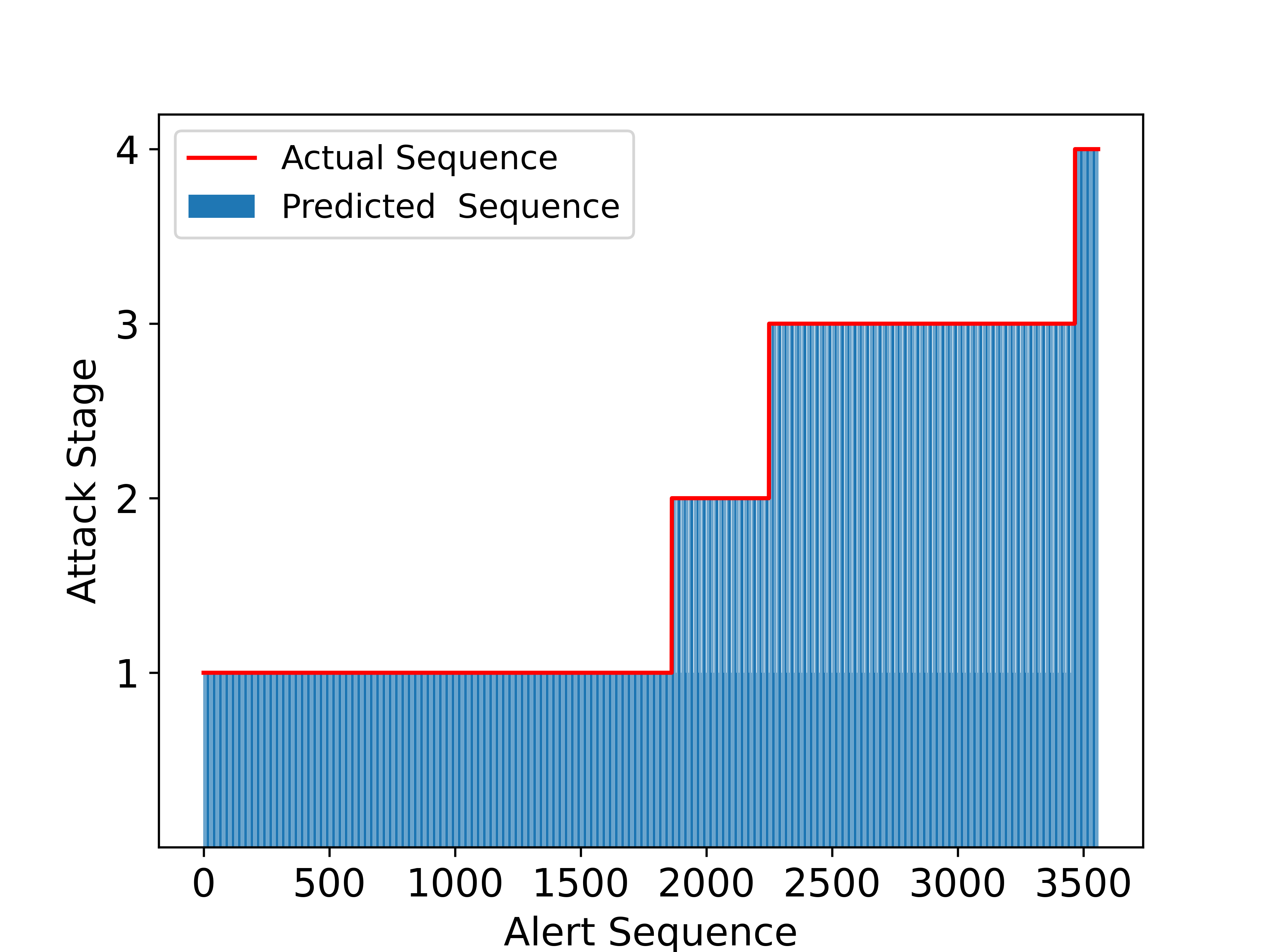}}
    \hfill
  \subfloat[\label{2exp1b}]{%
        \includegraphics[width=0.5\linewidth]{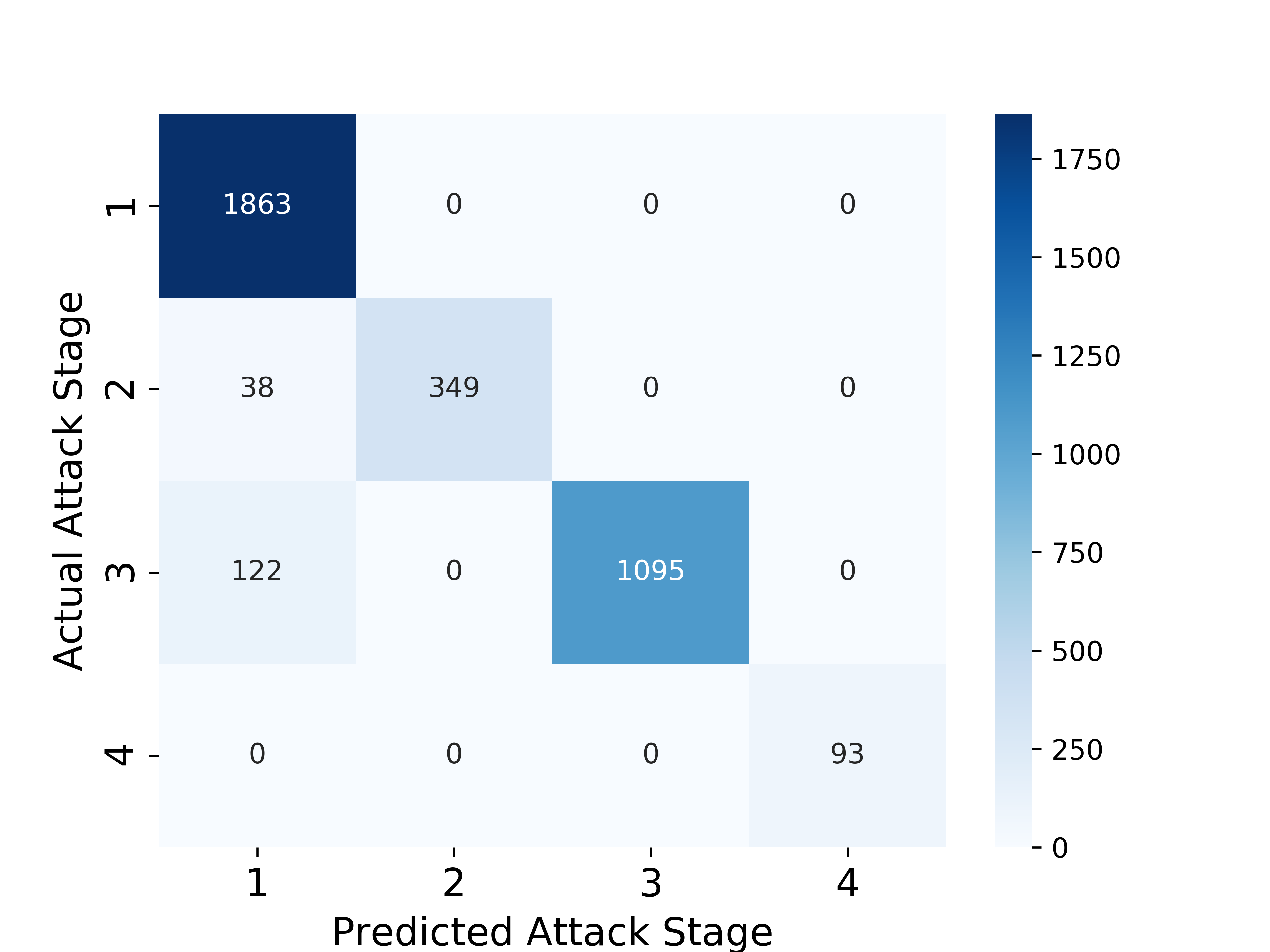}}
    \\
    \centering
  \subfloat[\label{2exp1c}]{%
       \includegraphics[width=0.5\linewidth]{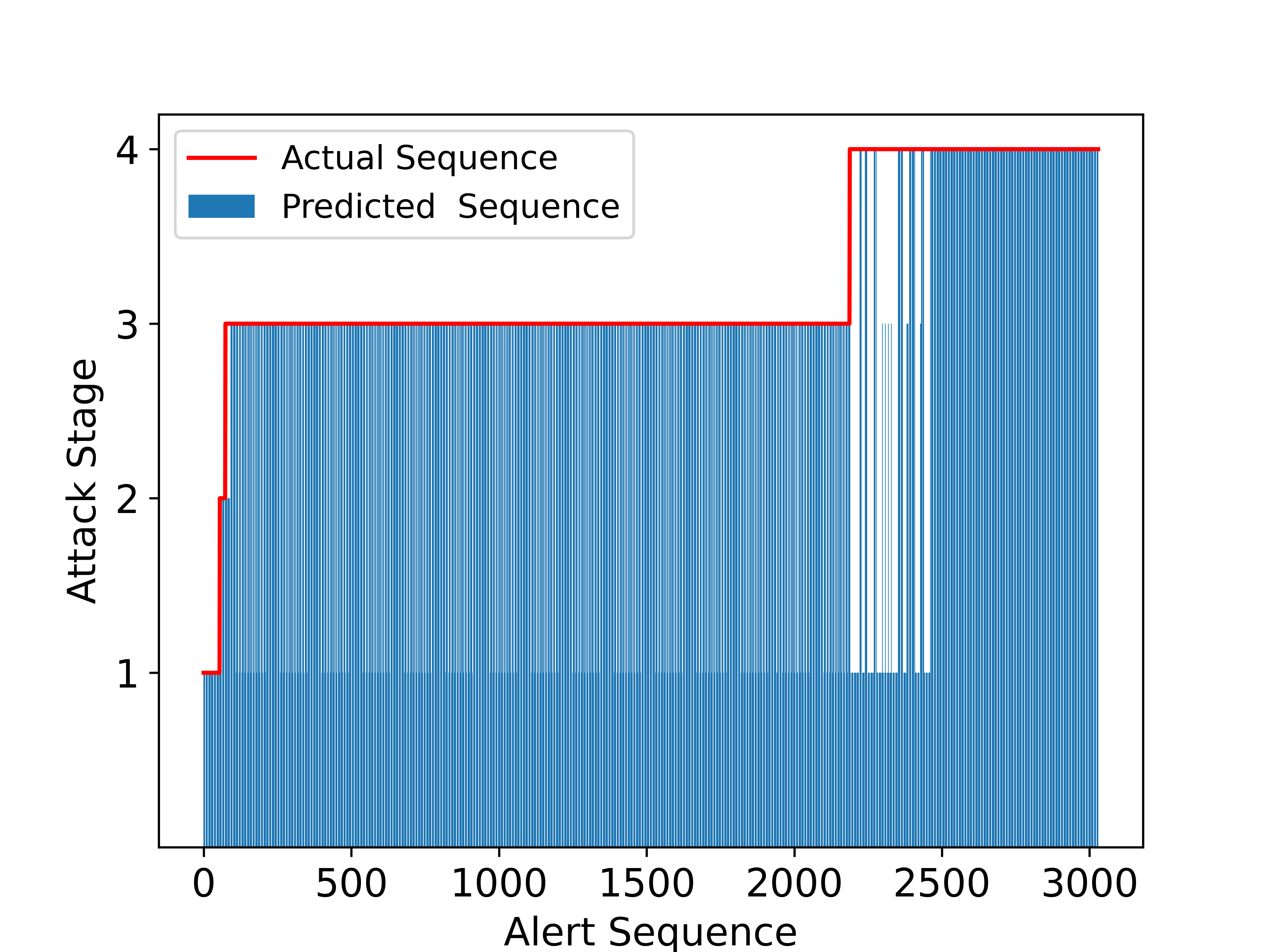}}
    \hfill
  \subfloat[\label{2exp1d}]{%
        \includegraphics[width=0.5\linewidth]{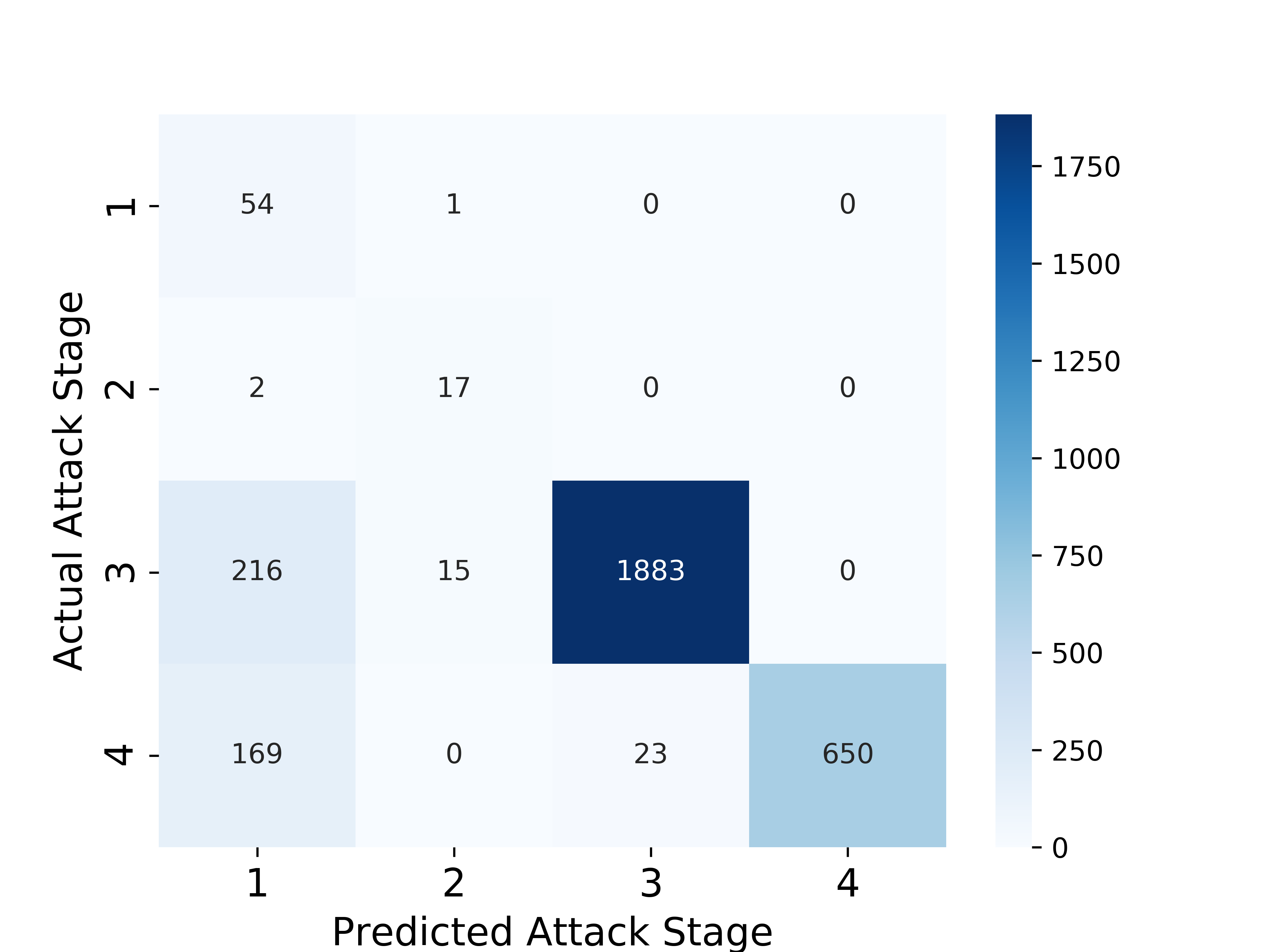}}
    \\
    \centering
  \subfloat[\label{2exp1e}]{%
       \includegraphics[width=0.5\linewidth]{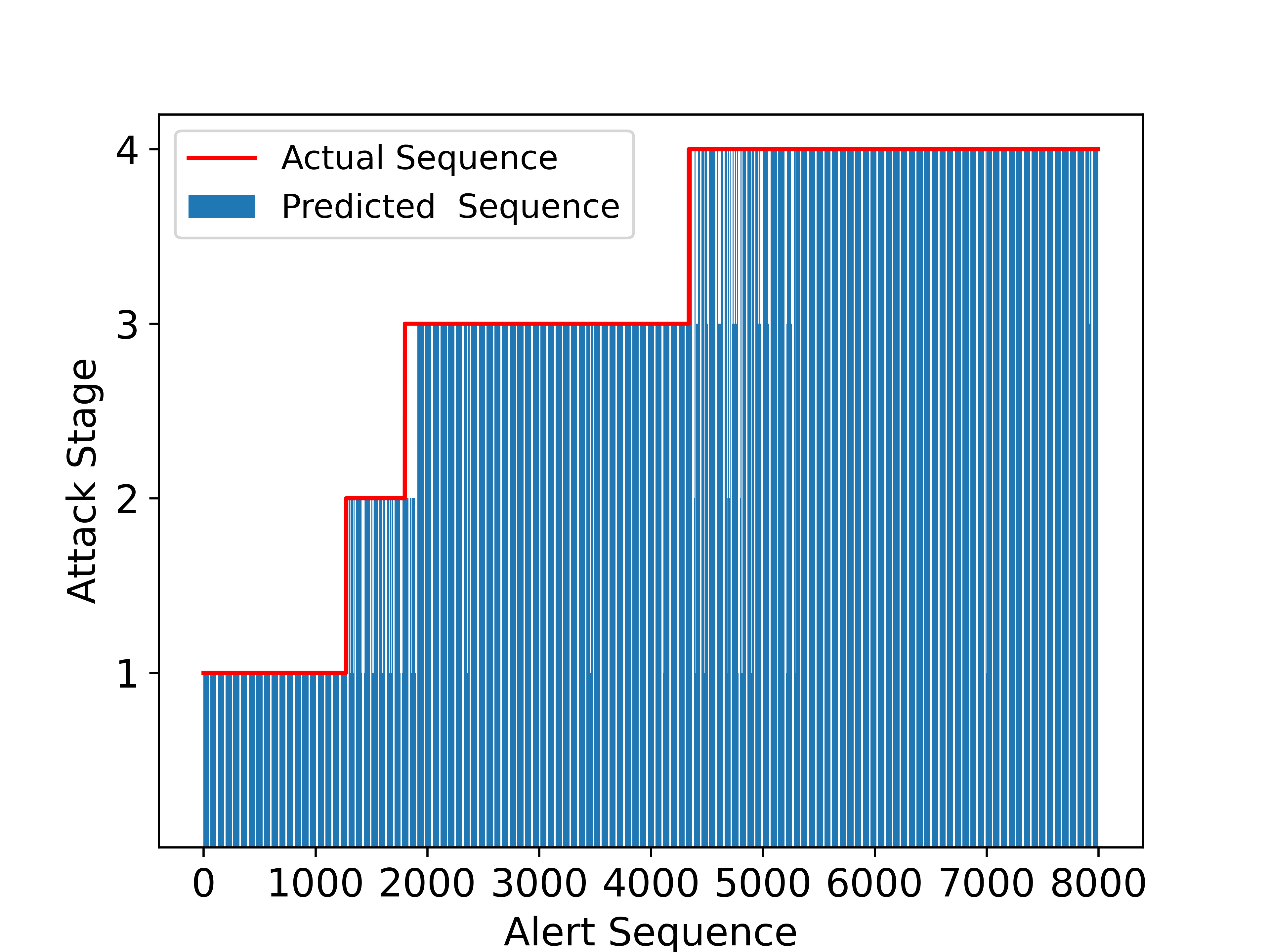}}
    \hfill
  \subfloat[\label{2exp1f}]{%
        \includegraphics[width=0.5\linewidth]{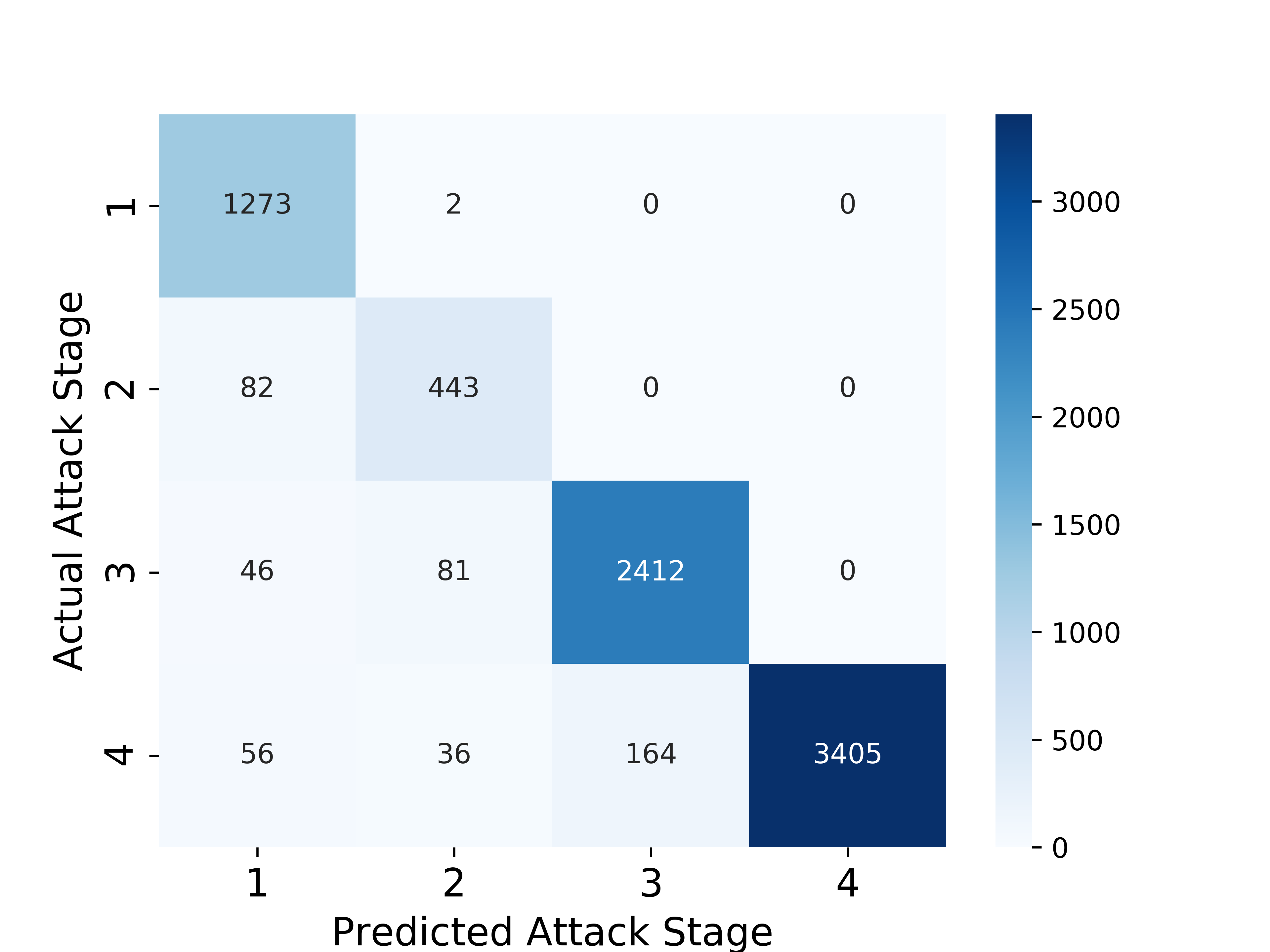}}
  \caption{Attack stage prediction and confusion matrix representation of prediction output for (a),(b) infiltration attack, (c),(d) HTTP DoS attack, and (e),(f) DDoS botnet attack.}
  \label{2exp1} 
\end{figure}

\begin{figure} 
    \centering
  \subfloat[\label{2exp2a}]{%
       \includegraphics[width=0.5\linewidth]{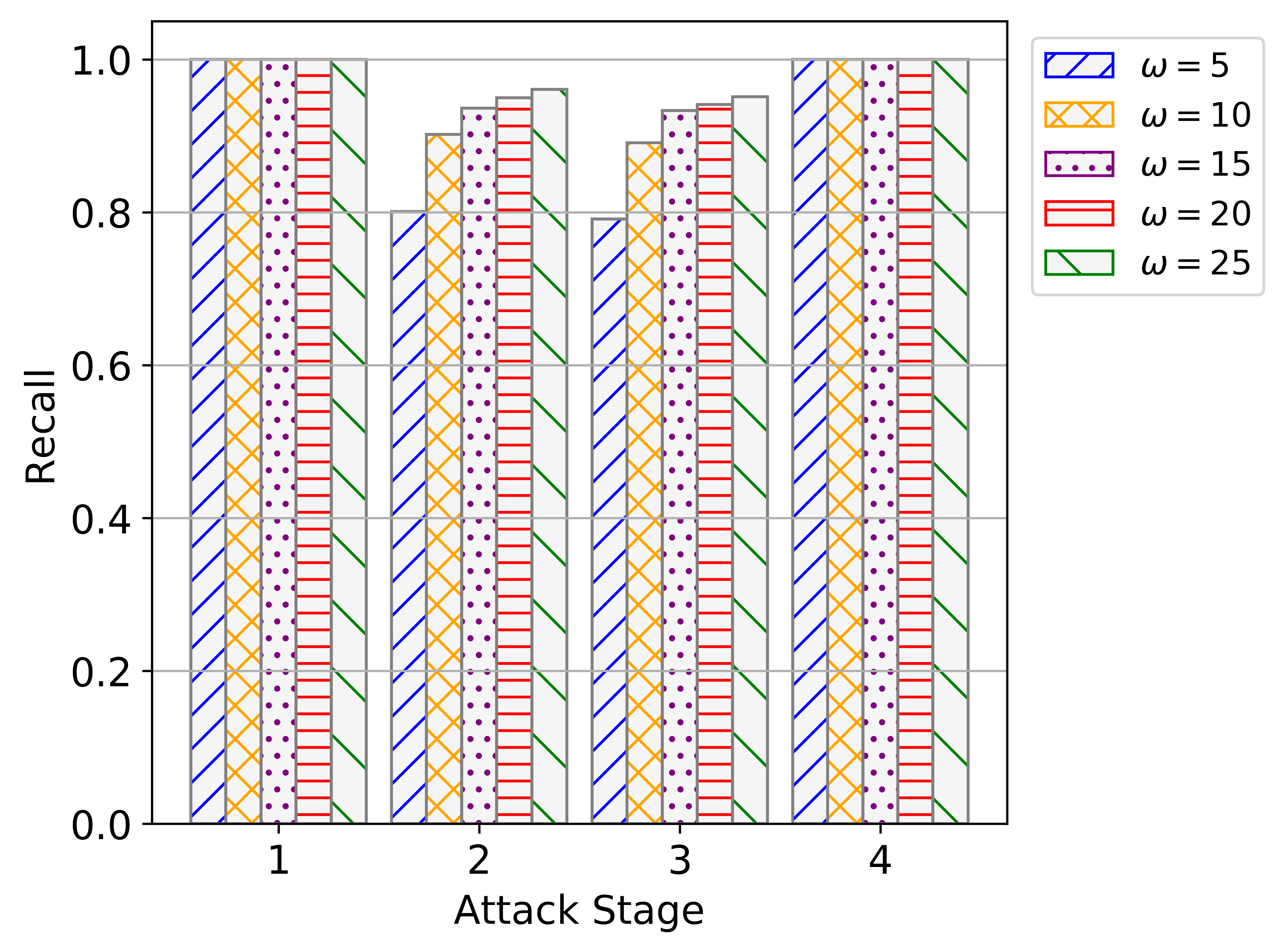}}
    \hfill
  \subfloat[\label{2exp2b}]{%
        \includegraphics[width=0.5\linewidth]{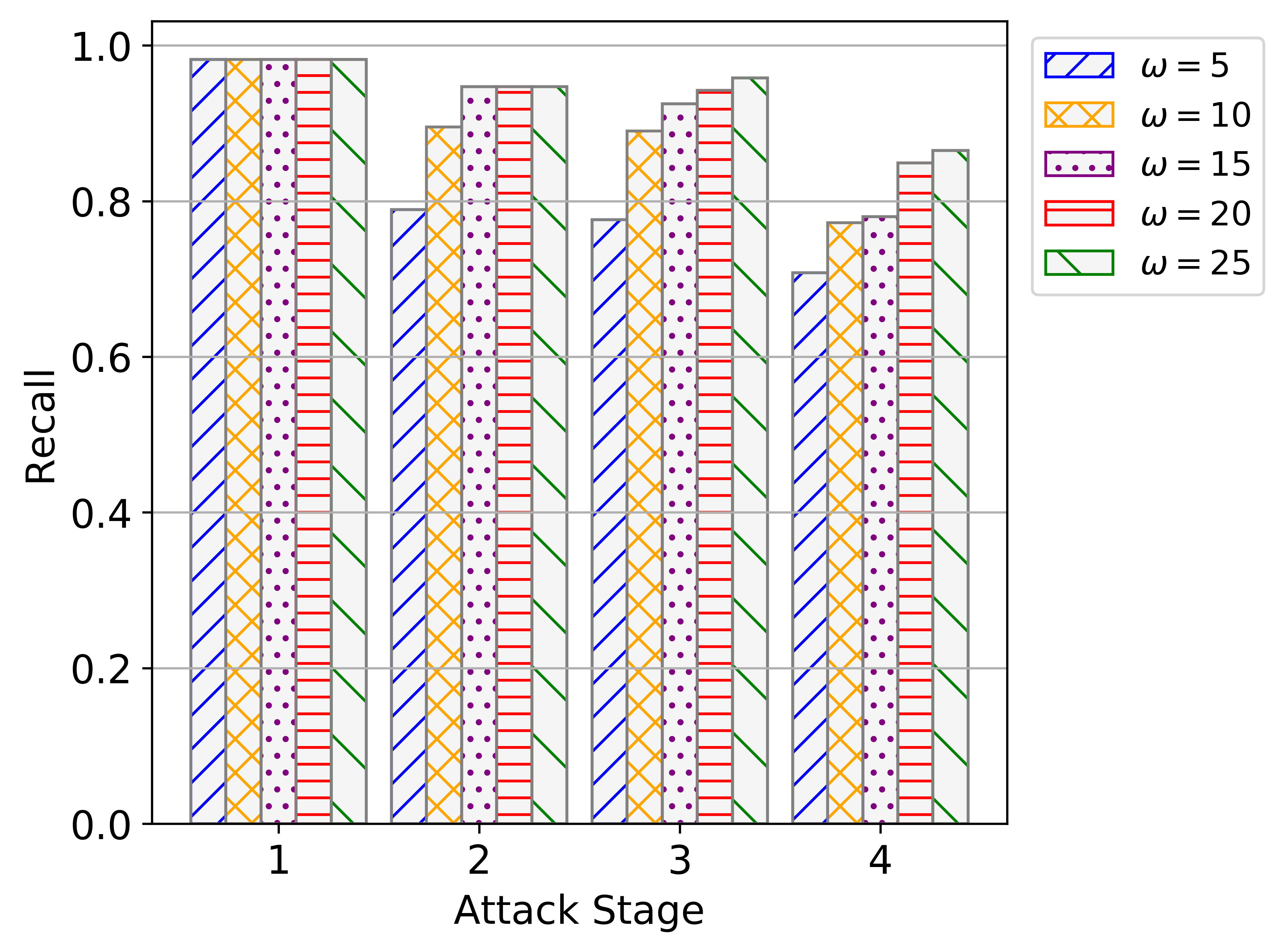}}
  \caption{Prediction recall for different alert sequence lengths ($\omega$) for (a) infiltration attack and (b) HTTP DoS attack.}
  \label{2exp2} 
\end{figure}

We have performed several experiments to determine the prediction efficacy of PRISM by varying the training data size and sampling ratios of the threat-aware sampling. All of the results presented here correspond to network traffic data sampled using threat-aware sampling at $\sigma=0.3$. The prediction output and the classification details of prediction output in the form of a confusion matrix is presented in Fig. \ref{2exp1}(a),(b) for infiltration attack, Fig. \ref{2exp1}(c),(d) for HTTP DoS attack, and Fig. \ref{2exp1}(e),(f) for DDoS botnet attack. It can be seen that the overall prediction performance is fairly decent for each attack type corresponding to $\sigma=0.3$ and $\omega = 10$. In the infiltration attack scenario, stages 1 and 4 of the attack are predicted without any error while stages 2 and 3 are incorrectly predicted as stage 1 for some instances. For HTTP DoS attack, most of the incorrect predictions are for attack stages 3 and 4. In the DDoS botnet attack scenario, incorrect predictions are mostly for attack stages 3 and 4 while there are few incorrect predictions in stage 2 of the attack as well. The prediction capabilities of PRISM are primarily dependant on the quality of data used in training the models. However, given a trained model, the runtime performance of the Prediction Engine is influenced by $\omega$, and generally longer length alert sequences are predicted by the model with more accuracy. But formation of long sequences entails increased waiting times to gather alerts, and in a realtime environment, the requirement is to process the alerts as quickly as possible for a timely prediction decision. Experiments are conducted to reveal the effect of different alert sequence sizes on the prediction correctness and the results are illustrated in Fig. \ref{2exp2}. To measure the prediction correctness, we have used recall that quantifies how well the model performs in identifying the actual attack stage in a given configuration. Fig. \ref{2exp2a} shows the prediction recall values for infiltration attack corresponding to different values of $\omega$. It can be seen that the prediction recall for stages 1 and 4 is ideal for all values of $\omega$, but for stages 2 and 3, the prediction recall values are less for lower values of $\omega$. Fig. \ref{2exp2b} illustrates the prediction recall for HTTP Dos attack and it can be seen that the prediction recall improves with the increase in $\omega$ for stages 2, 3 and 4. If Fig. \ref{2exp1}(b),(d) is observed together with Fig. \ref{2exp2}(a),(b), then it can be realised that lower values of $\omega$ exacerbate the error in prediction, especially for attack stages that are prone to be predicted incorrectly.     

\begin{figure} 
    \centering
  \subfloat[\label{2exp3a}]{%
       \includegraphics[width=0.5\linewidth]{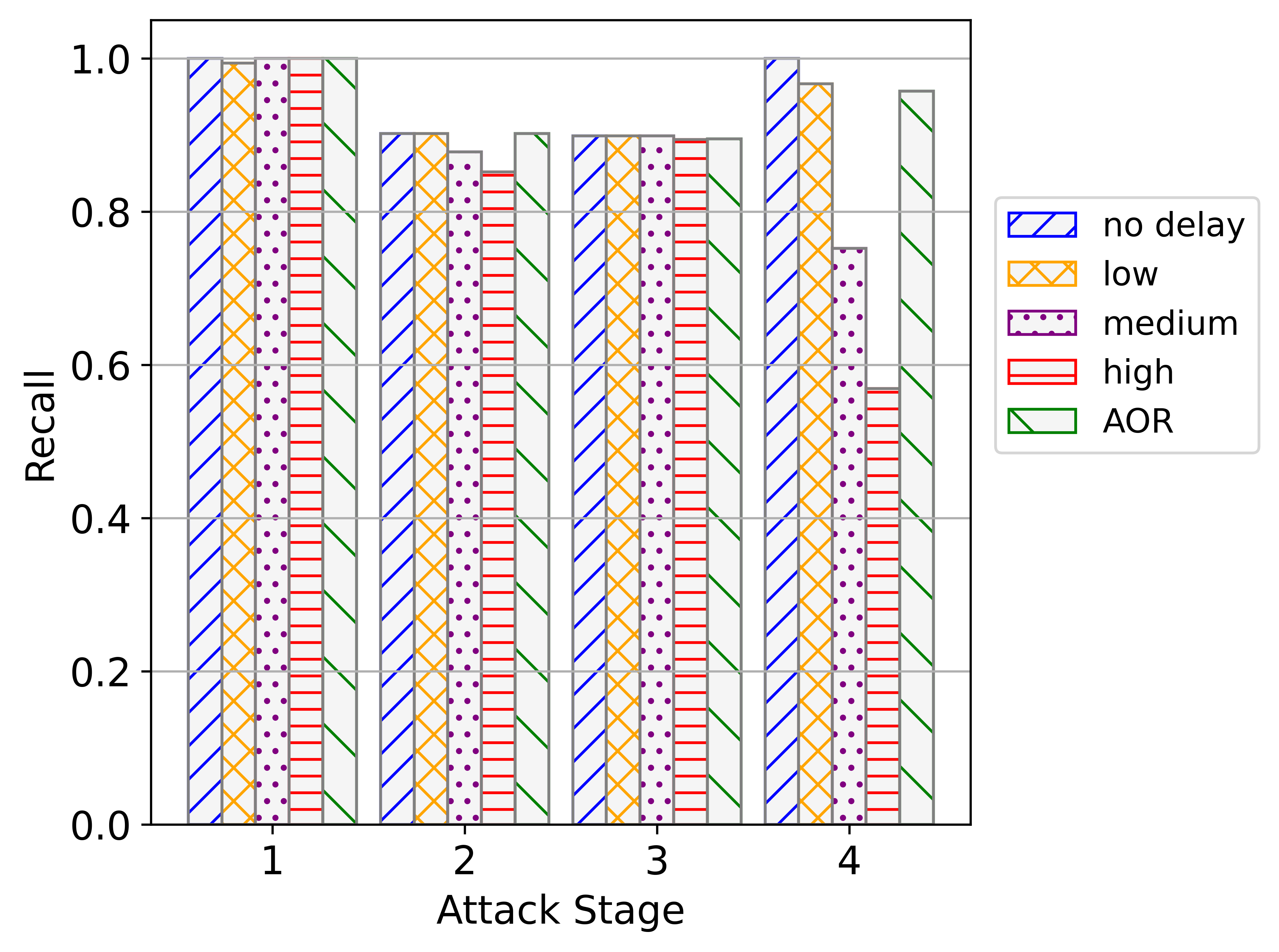}}
    \hfill
  \subfloat[\label{2exp3b}]{%
        \includegraphics[width=0.5\linewidth]{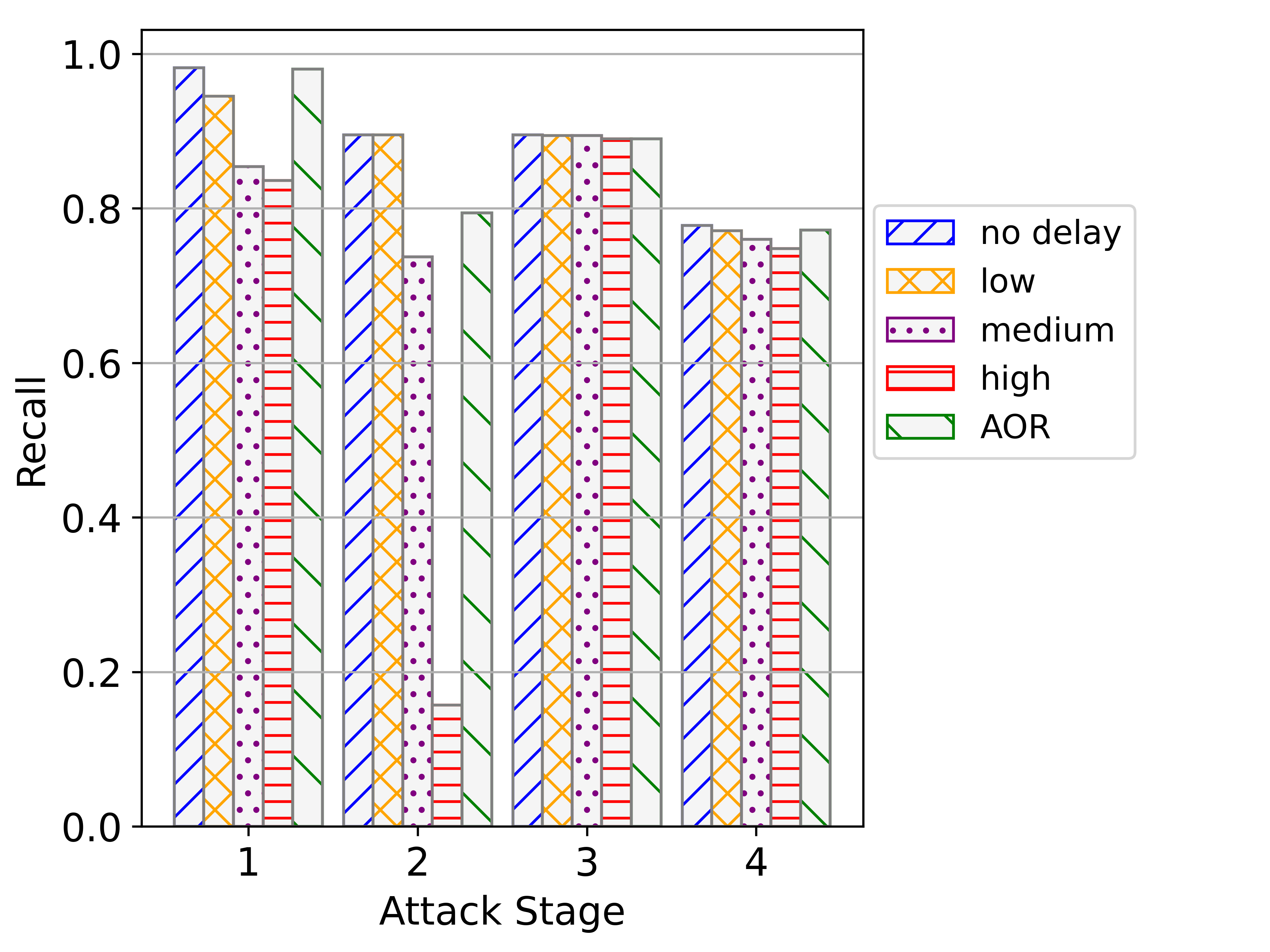}}
  \caption{(a),(b) Prediction recall corresponding to different delay configurations for infiltration and HTTP DoS attacks, respectively.}
  \label{2exp3} 
\end{figure}

To ascertain the effect of alert stream management on the prediction performance we have engineered distributed alert reporting mechanism as discussed in Section 4. The experimentation has been carried out with $\omega=10$ using the three delay configurations: low, medium and high where the value of $\mu_{max}$ is set to be 30\% lower than the value of $\Delta_{max}$, as detailed in Table \ref{distAlert}. The prediction recall corresponding to the three delay configurations and the alert sequencing using the AOR procedure on the high delay configuration impacted alerts is shown in Fig. \ref{2exp3}. It can be seen that as the distributed alert reporting latency becomes more random, that is, with higher alert delay configuration, the ability of the model to predict attack stages correctly is significantly reduced, especially for those attack stages that have fewer number of alerts. This phenomenon can be observed through the prediction recall values for high delay configuration at stage 4 of the infiltration attack in Fig. \ref{2exp3}a and stage 2 of the HTTP DoS attack in Fig. \ref{2exp3}b. The effectiveness of alert stream management is also apparent as the AOR procedure even with $\mu_{max}$ estimate being 30\% less than $\Delta_{max}$ value is able to restore the original order of the alerts significantly well. It is to be noted that there is a cost associated to the application of AOR procedure that every time an alert forwarding window is formed, there is a wait of $\mu_{max}$ before the alerts are sent to the Prediction Engine.   

Prediction recall is one aspect to quantify the prediction system's performance. In a realtime environment, it is necessary to measure how soon the prediction operation is able to detect attack stage transitions during the progression of an attack. The sooner the prediction apparatus is able to identify the change in attack stage, the more effective response to thwart the attack can be formulated. To measure the ability of the Prediction Engine to detect the correct attack stage as soon as the attack transitions from one stage to another, the metric Early Detection Index (EDI) is proposed and is expressed in Eq. \ref{edi}.          

\begin{equation}
    EDI = \frac{\sum_{i=1}^N \frac{o_i^{l} - o_i^{f}}{C_i} \: \varepsilon_i}{H}
\label{edi}
\end{equation}

\begin{figure} 
    \centering
  \subfloat[\label{2exp4a}]{%
       \includegraphics[width=0.5\linewidth]{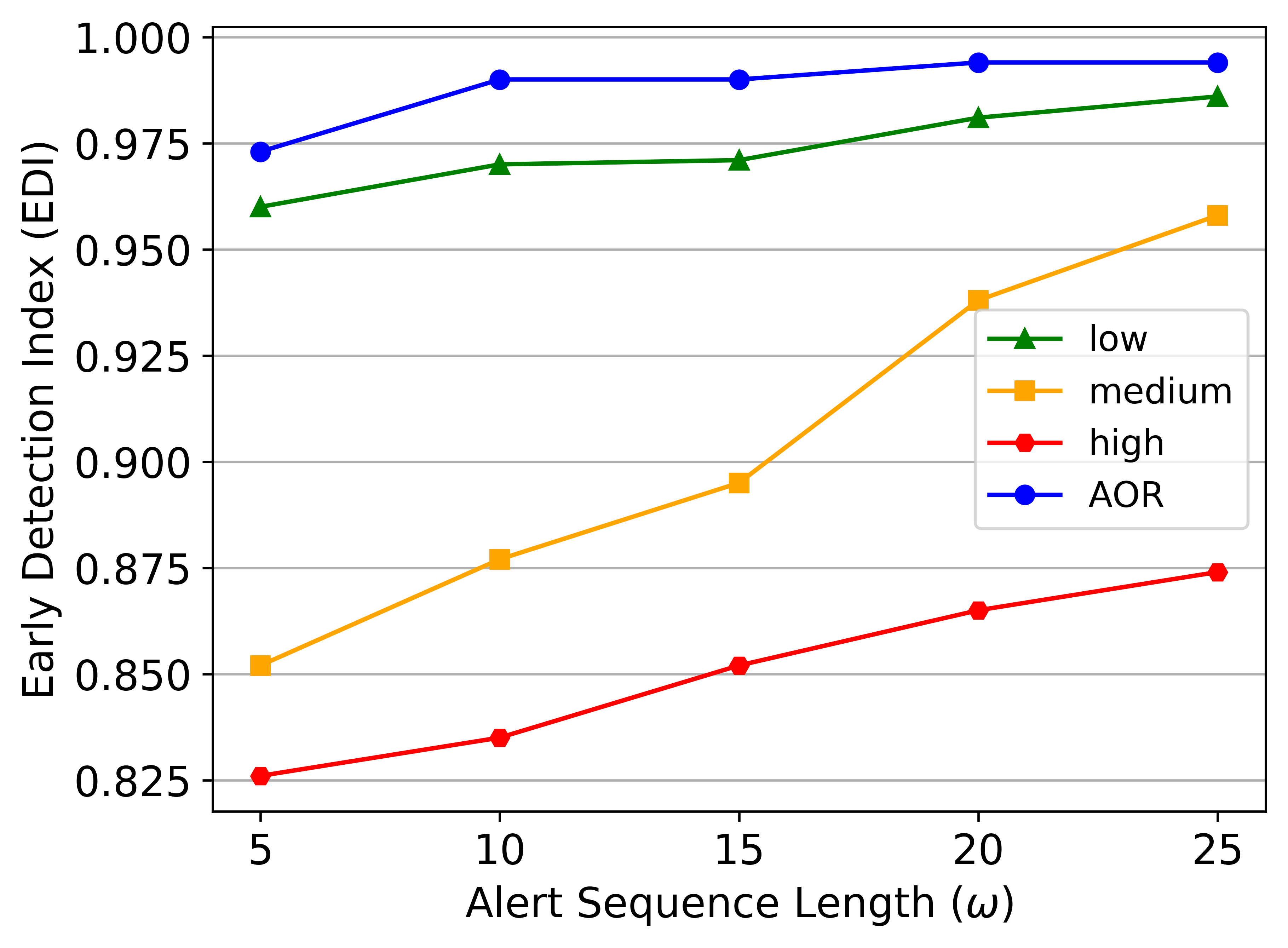}}
    \hfill
  \subfloat[\label{2exp4b}]{%
        \includegraphics[width=0.5\linewidth]{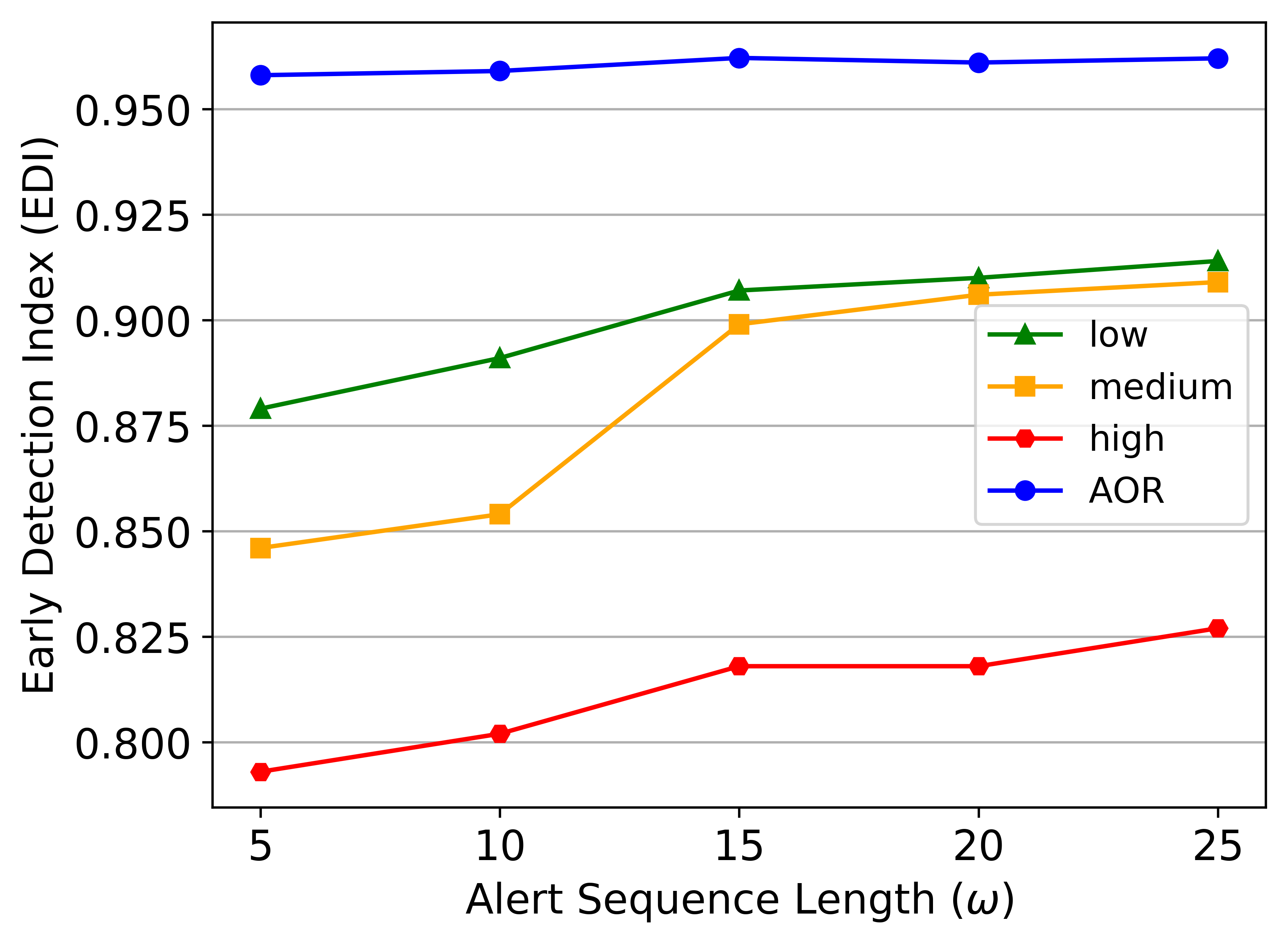}}
  \caption{EDI corresponding to different delay configurations for (a) infiltration attack and (b) HTTP DoS attack.}
  \label{2exp4} 
\end{figure}

Where $o_i^l$ represents the alert number of the last alert in attack stage $i$ in original labeled data (ground truth) and $o_i^f$ is the alert number of the first alert that the model predicts to be of attack stage $i$, that is, the alert that detects the beginning of attack stage $i$. $C_i$ is the total number of alerts corresponding to attack stage $i$ in the original labelled data and $\varepsilon_i$, is the numerical weight corresponding to the risk associated with attack stage $i$. Any $\varepsilon$ value can be assigned to an attack stage with advanced attack stages being given higher values and in our experiments the attack stage number itself is used as the $\varepsilon$ value. $H$ is the sum of attack stage risk values, that is, $H = \sum_{i=1}^N \varepsilon_i$. Fig. \ref{2exp4}(a),(b) shows the EDI with respect to different values of $\omega$ for the three delay configurations and alert order correction using the AOR procedure. It can be observed that the EDI for both of the attacks is quite less in the case of high delay configuration as compared to medium and low delay configurations. Whereas, the EDI is near the maximum value of 1 in the case of alert stream being subjected to the AOR procedure. This demonstrates that the use of alert stream management not only benefits the prediction correctness but also significantly improves the ability of the Prediction Engine to detect attack stage transitions rapidly. 


\begin{figure*}
    \centering
    \subfloat[System Availability]{\includegraphics[width=.34\textwidth]{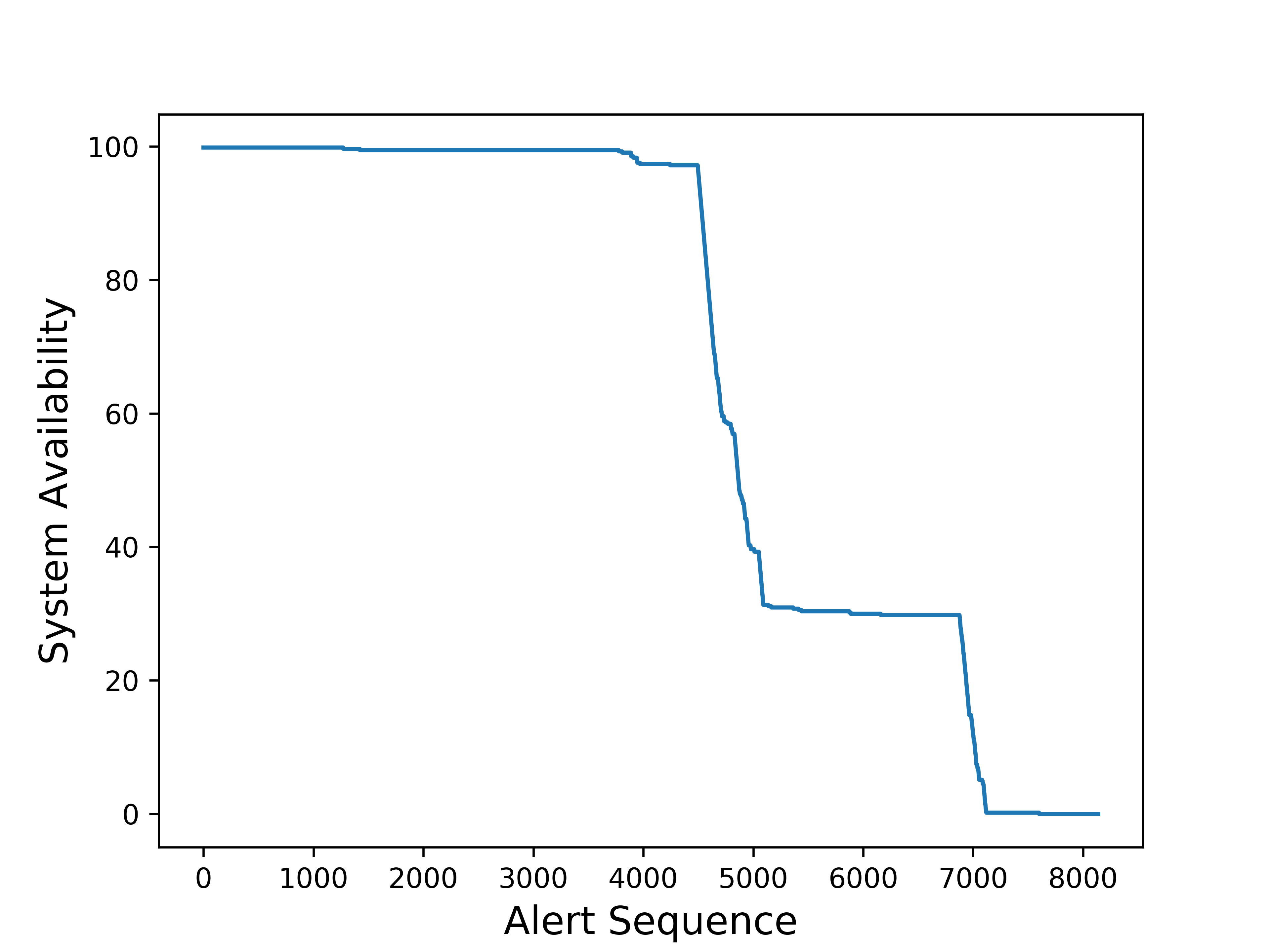}}
    \subfloat[Threat Perceptivity]{\includegraphics[width=.34\textwidth]{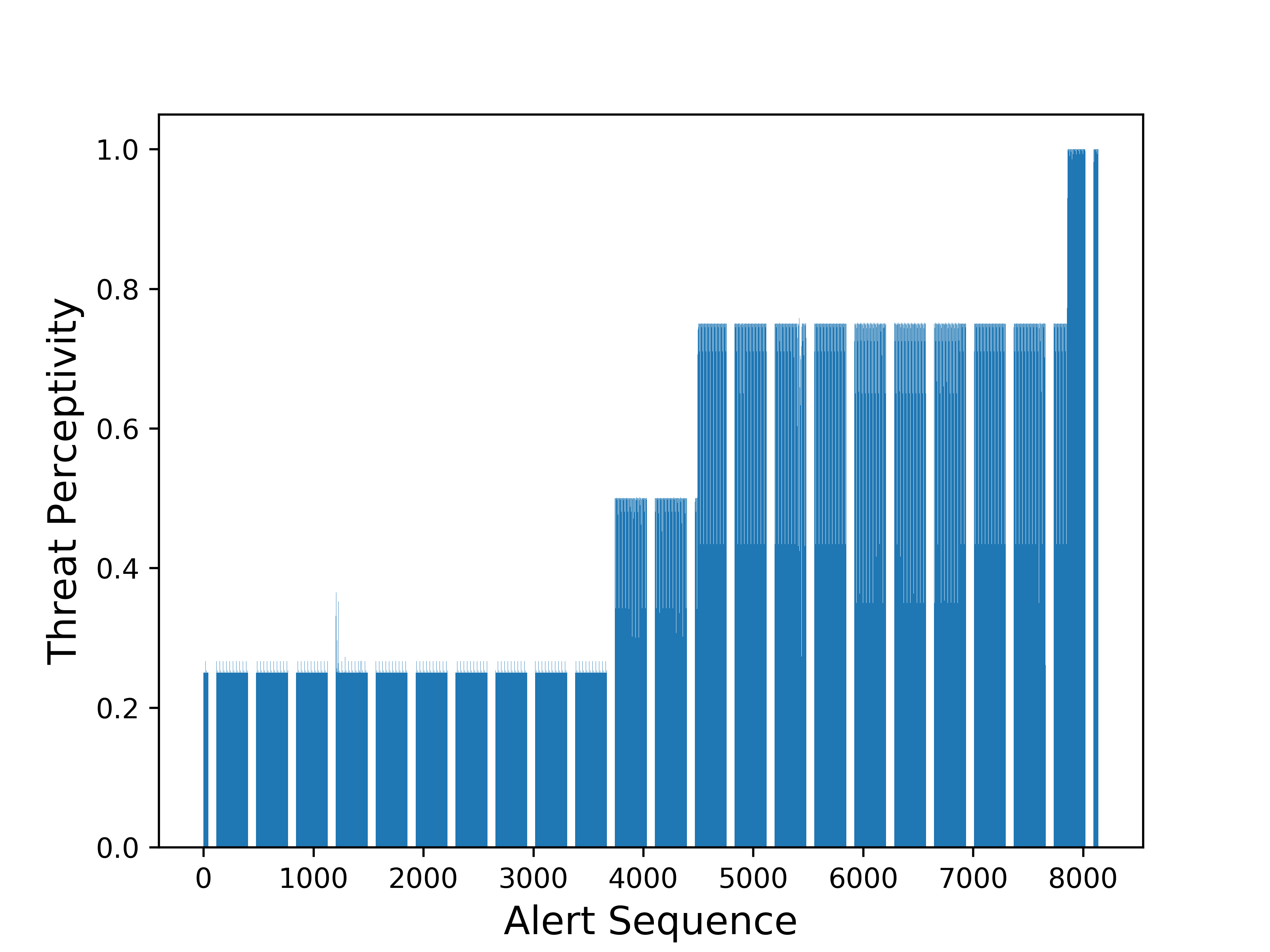}}
    \subfloat[System Degradability]{\includegraphics[width=.34\textwidth]{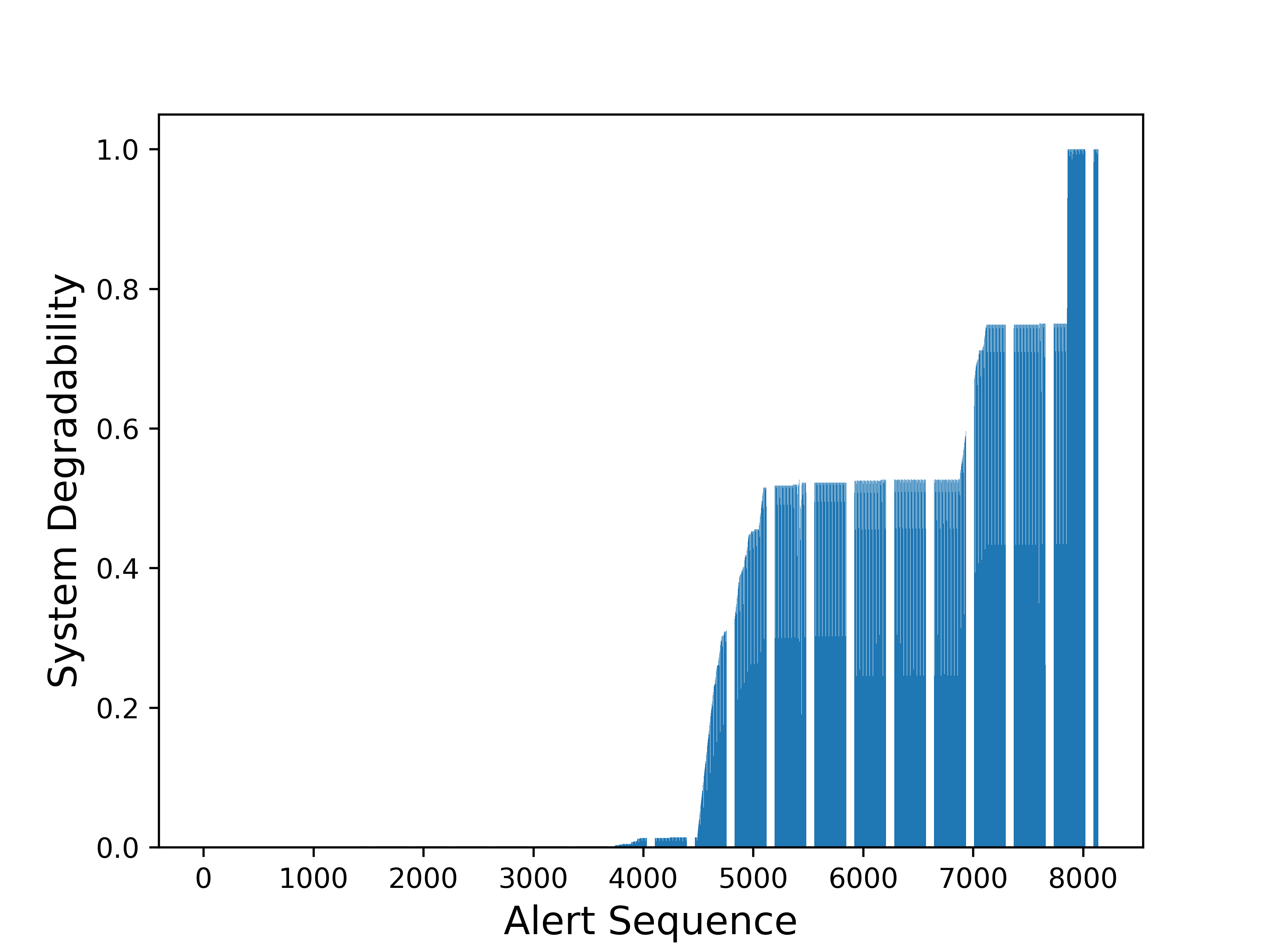}}
    \caption{Security Analyzer output during infiltration attack scenario.}
    \label{fig:sysanalyzer}
\end{figure*}

\begin{figure*}
    \centering
    \subfloat[System Availability]{\includegraphics[width=.34\textwidth]{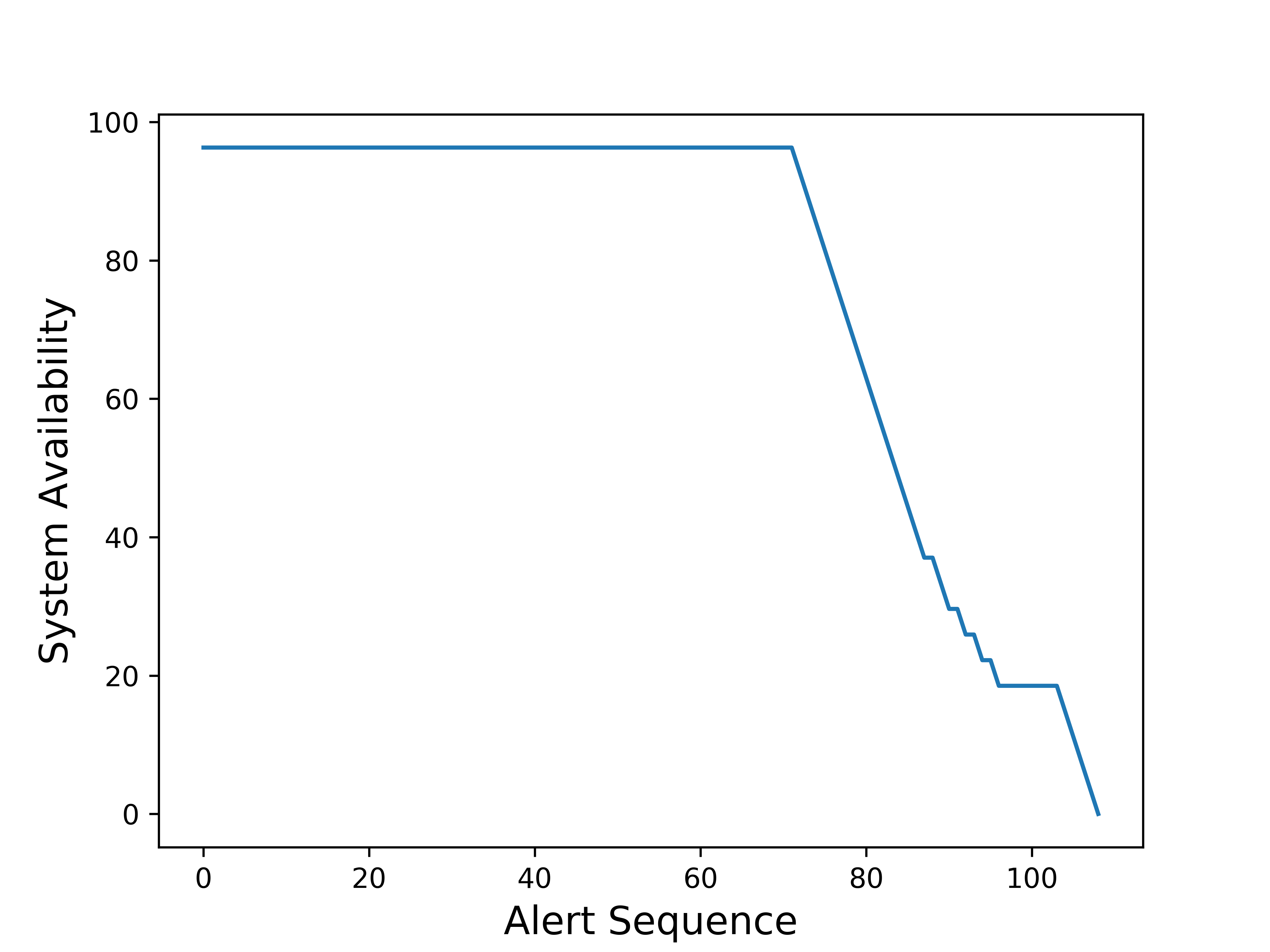}}
    \subfloat[Threat Perceptivity]{\includegraphics[width=.34\textwidth]{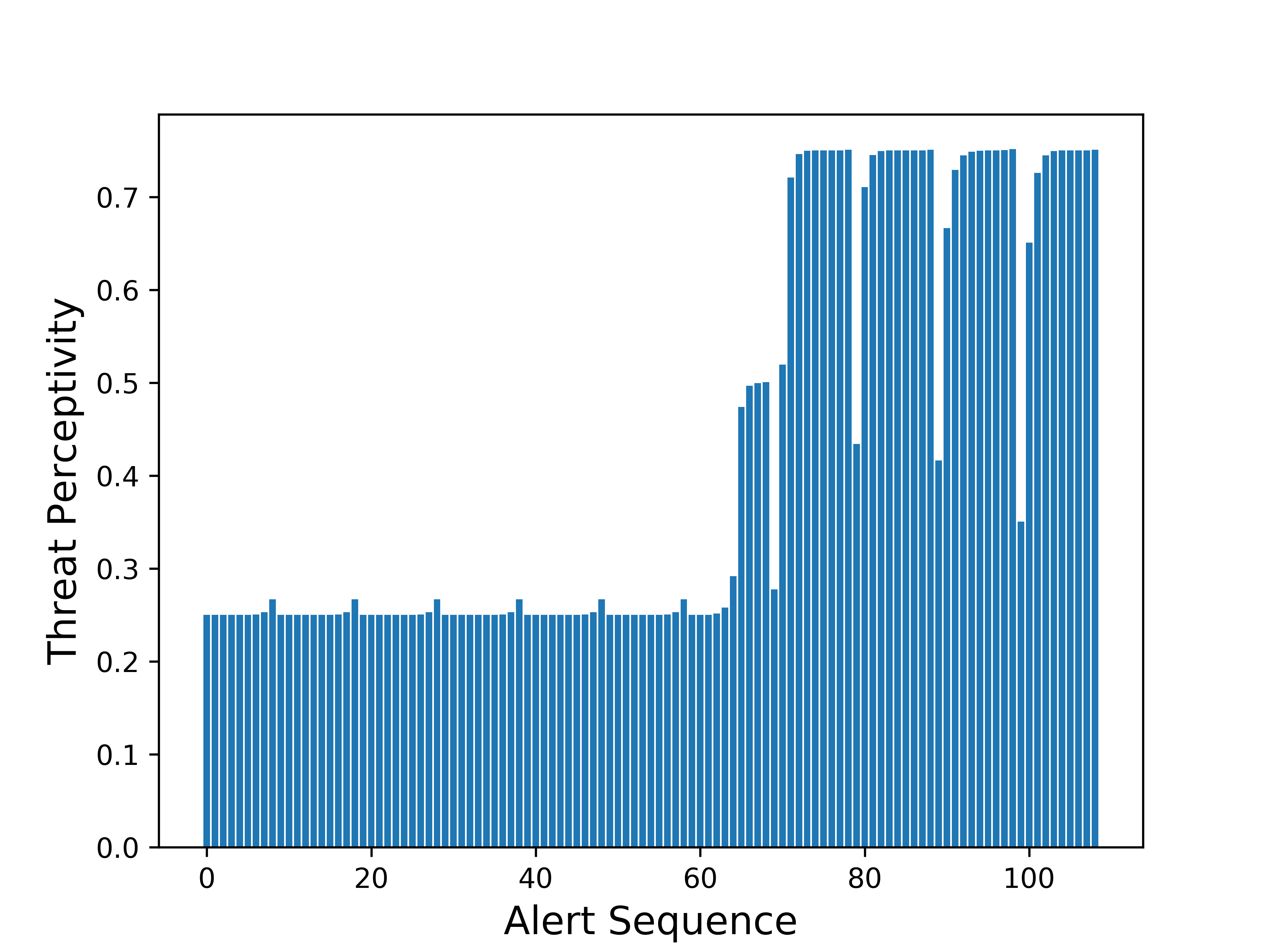}}
    \subfloat[System Degradability]{\includegraphics[width=.34\textwidth]{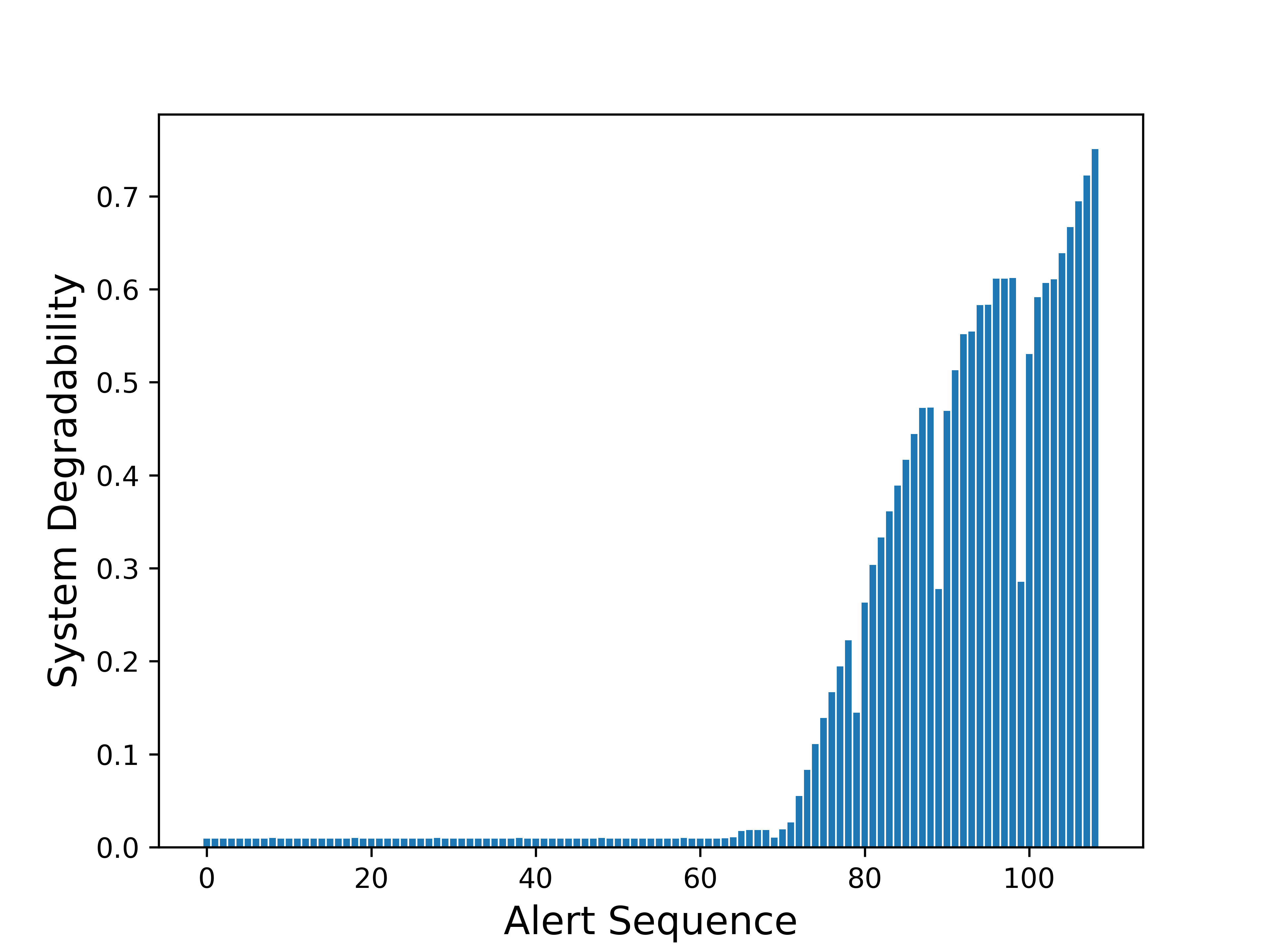}}
    \caption{Security Analyzer output corresponding to 1\% of sampled traffic.}
    \label{fig:sysanalyzer_1}
\end{figure*}

\subsection{Security Analysis for Cyber Situational Awareness}
We have evaluated the functionality of Security Analyzer on the alerts generated from infiltration attack data. In the experiments the network traffic is not sampled and the value of $\omega$ is set to be 10. Also, the alerts from different surveillance zones are not subjected to any additional delay. The metrics are computed and visualized by the Security Analyzer as each of the alert is received in realtime, but here we present the plots of the complete attack. In Fig \ref{fig:sysanalyzer}a, system availability of the network is shown with respect to the alerts. It can be seen that there is little effect on the system availability during the first half of the attack, but during the second half the system availability reduces drastically. Threat perceptivity corresponding to the stream of incoming alerts is illustrated in Fig. \ref{fig:sysanalyzer}b. As expected, the threat perceptivity increases as the attack progresses into advanced stages. In Fig. \ref{fig:sysanalyzer}c, system degradability for the case of infiltration attack is presented, and it can be observed that as the attack progresses, with the decreasing system availability, the system degradability increases. Fig. \ref{fig:sysanalyzer_1} presents the output of Security Analyzer corresponding to infiltration attack scenario with just using 1\% of the network traffic that is sampled using the threat-aware sampling. In addition to exhibiting the effect of a very high sampling rate, this experimental setting reflects the impact of a fast moving stealthy attack that achieves the target by generating a minimal amount of alerts. It can be observed that the system availability, threat perceptivity and system degradability changes drastically to the worst levels within the last 30 reported alerts. This shows the limitation of Security Analyzer in assisting human-driven response operations and reaffirms the proposition that automated response mechanisms are better suited to foil such attack types as discussed in \cite{javed}.

\section{Related Work}
Many IDS architectures have been proposed over the years with a certain IDS architecture being dominant for a particular network type. For enterprise networks a centralized IDS deployment is the most common in which the IDS scrutinises all traffic passing through the external routers that connect the enterprise network to the internet \cite{zeeshan}. Internet of Things (IoT) networks are generally protected by distributed IDS as there is no central location from where data of the whole network can be monitored \cite{chaabouni}. A distributed IDS architecture for enterprise networks is proposed in \cite{LiMIT} that has IDS components distributed into different regions of the network. Each regional detector reports malicious activities of its sphere to a global detector that uses Sequential Hypothesis Testing (SHT) to determine whether the whole enterprise network is under attack. In \cite{hardegen}, a fully distributed architecture is proposed in which the IDS resources are divided in three layers: in-network, local and global. The in-network IDS instances directly observe network traffic and send network anomaly detection alerts to local IDS instances. Global IDS instances perform analysis on the information received from local IDS instances to determine the overall intrusion status. In \cite{bovenzi}, a two-stage hierarchical intrusion detection approach is presented that identifies network traffic anomalies in the first stage and in the second stage classifies the detected anomalies into different attacks. In \cite{hamids}, a hierarchical network traffic monitoring approach is presented that uses Bro IDS \cite{zeek} at different layers to detect network-based attacks. A collaborative anomaly detection framework has been proposed in \cite{xie} that uses network virtualization techniques to implement a global situation detection and local behavior detection model. The framework uses Hidden Markov Random Field (HMRF) to model network behavior and spatial Markovianity to model the network behavior's spatial context. Recently, the use of blockchain technology in distributed IDS design has become popular. In \cite{alkadi}, a distributed IDS has been proposed that uses blockchain and smart contracts to mitigate threats on collaborating IDSs on different cloud networks. None of the abovementioned distributed IDSs, unlike PRISM, have the capability to identify multi-stage attacks and predict the attack progression.   


\par Several intrusion prediction, attack projection, attack intention recognition and network security situation forecasting techniques have been proposed \cite{husak}. In this paper we will only discuss intrusion detection and prediction using machine learning models, specifically HMM-based intrusion prediction. In \cite{fu} a machine learning-based malicious traffic detection system is proposed that identifies anomalous network traffic by implementing a statistical clustering-based model on the frequency domain features of network traffic. Similarly, in \cite{mirsky} an autoencoder driven anomalous network traffic detection system is proposed that models the normal network traffic behavior and identifies deviations from the normal behavior as anomalies. Though these solutions can detect zero-day attacks as behavior anomalies, but the binary output they produce as normal or malicious for the observed traffic pattern is of little use to the security analysts for response operations. Moreover, such solutions do not have the ability to predict attack progression of multi-stage attacks. PRISM on the other hand can detect complex multi-stage attacks with the ability to identify attack progression stages in realtime. The use of HMMs for network intrusion detection is not new, in \cite{cho, joshi, larrohando, ariu, gornitz} anomaly detection systems using HMMs are presented. Recently, HMMs have been extensively used for network intrusion prediction. An intrusion prediction and prevention system has been presented in \cite{haslum} that uses HMM to predict the next step of an attacker. The intrusion prediction information is then used for intrusion prevention operations according to a set of predetermined rules. In \cite{farhadi} data mining driven algorithm is proposed that mines the stream of IDS alerts for attack scenario extraction. An HMM-based alert correlation system is then introduced that predicts the next class of attacks to be launched by the intruder. Frameworks to predict multi-step network attacks using HMMs are proposed in \cite{sendi, holgado}. In \cite{shawly}, HMM-based architectures have been proposed that predict the attack stage when several multi-stage attacks are in action simultaneously. Lately, to address the issue of limited availability of network intrusion detection datasets, a transfer learning-based approach is proposed that trains the HMM model using a labelled dataset and then adapts the model to a target unlabelled dataset \cite{chadza}. The aforementioned intrusion prediction techniques only offer a centralized approach to process network traffic. PRISM on the other hand presents an intrusion prediction architecture with distributed traffic processing. Moreover, PRISM addresses the challenges associated with the distributed design by introducing an efficient alert stream management mechanism.

\par The impact of sampling techniques designed for network traffic engineering on the performance of network intrusion detection is well studied. In \cite{mai}, the suitability of four commonly used sampling techniques for intrusion detection applications has been investigated. The intrusion detection performance of random packet sampling, random flow sampling, sample-and-hold and smart sampling is compared. It has been shown that flow-based sampling performs better than packet-based sampling. Both sample-and-hold and smart sampling are found to be biased towards large volume flows, and are not suitable to detect several attack types. In \cite{androulidakis}, smart sampling and selective sampling techniques are evaluated for intrusion detection applications. Results demonstrate that selective sampling does not exhibit much bias towards large volume flows unlike smart sampling. In \cite{bartos}, a sampling scheme designed for network intrusion detection has been proposed. The developed sampling mechanism is based on late and adaptive sampling process. The late sampling process extracts flow features and calculate flow statistics. Adaptive sampling process then selects outlier flows with more preference as compared to flows with similar flow statistics. Recently, machine learning-based sampling techniques have been employed for network intrusion detection applications. In \cite{su}, a hierarchical clustering-based sampling technique is proposed that uses structural and temporal features of suspicious flows in a two-step hierarchical clustering algorithm. A trust-based intrusion detection mechanism is proposed in \cite{meng} that uses traffic filtering and sampling to reduce the volume of network traffic received at the IDS. Traffic filtering is performed by blacklisting the nodes that have been observed to emit malicious packets and traffic sampling is accomplished using the systematic sampling approach that is similar to n-out-of-N random sampling technique. In \cite{sibai}, the impact of sampling on anomaly detection is studied by comparing the performance of four sampling techniques. Weighted Round Sampling (WRS), deterministic sampling, Simple Random Sampling (SRS) and chain-sampling algorithms are evaluated. It has been shown that WRS performs better than the other evaluated sampling techniques. PRISM introduces a novel threat-aware sampling scheme that distinguishes from the existing sampling techniques by employing an attacker behavior model-based ranking mechanism that uses the vulnerability and inter-device reachability information to sample more traffic from the devices that are more likely to be targeted by the attacker.

\section{Conclusion}
In this paper, we present PRISM, a hierarchical intrusion detection architecture that can predict the progression of complex multi-stage attacks in realtime by processing a fraction of the total network traffic. PRISM employs an attacker behavior model-based threat-aware sampling scheme and a distributed network traffic monitoring mechanism to streamline its traffic processing operations. Due to its distributed structure, PRISM faces the challenge of alert order preservation that introduces errors in the prediction process. PRISM addresses this challenge by implementing an efficient alert stream management mechanism. Extensive experimentation has been conducted to evaluate the performance of PRISM under different conditions. It has been shown that PRISM reduces the network traffic processing overhead by up to 750\% as compared to a standard IDS while being able to predict the attack progression accurately. PRISM also demonstrates the ability to predict progression of an attack to advanced stages as soon as the transitions unfold, providing the response infrastructure additional margin to operate in a time constrained environment. Furthermore, multiple security metrics have been proposed that showcase a holistic security outlook of the system to support intrusion response operations.


%




\appendices
\section{Sequence Recognition and Decoding Using HMMs}
\subsection{Forward Algorithm}
Forward algorithm uses the forward variable $\zeta_t(i) = P(o_1,o_2,..,o_T,q_t = s_i|\lambda)$ which is the probability of the observation sequence $O = \{o_1,o_2,..,o_T\}$, and state at time $t$ being $s_i$, given the model $\lambda$. Forward algorithm solves for $\zeta_t(i)$ inductively, using the initialization, induction and termination steps illustrated in Eqs. \ref{forward1}, \ref{forward2} and \ref{forward3}, respectively \cite{HMMRabiner, HMMIntro}.

\begin{equation}
    \zeta_1(i) = \pi_i b_i (o_1), \;\;\;\;\;\;\; i \in [1, N] 
    \label{forward1}
    \tag{A.1}
\end{equation}
\begin{equation}
\begin{split}
    \zeta_{t+1}(j) = \sum_{i=1}^{N}\zeta_t(i)a_{ij}b_j(o_{t+1}), \;\;\;\;\;\;\; & j \in [1, N] \\
    & t \in [1, T-1]
\end{split}
\label{forward2}
\tag{A.2}
\end{equation}
\begin{equation}
    P(O|\lambda) = \sum_{i=1}^{N}\zeta_T(i)
\label{forward3}
\tag{A.3}
\end{equation}

\subsection{Viterbi Algorithm}
Viterbi algorithm takes a holistic approach in solving for the most likely state sequence $Q = \{q_1, q_2, .., q_T\}$ corresponding to the alert observation sequence $O = \{o_1, o_2, .., o_T\}$. The variable $\nu_t$, expressed in Eq. \ref{viterbi1}, navigates through the solution \cite{HMMRabiner, HMMIntro}.

\begin{equation}
    \nu_t(i) = \underset{q_1, q_2, .., q_{t-1}}{\max} P(q_1, .., q_{t-1},o_1, .., o_t, q_t = i | \lambda)
\label{viterbi1}
\tag{A.4}
\end{equation}

Where, $\nu_t(i)$ is the maximum probability for a single state sequence accounting for $t$ observations and $s_i$ being the last state in the state sequence. Using induction, Eq. \ref{viterbi2} can be derived.

\begin{equation}
    \nu_{t+1}(j) = \underset{i = 1, ..,N}{\max} \nu_t(i)a_{ij}b_j(o_{t+1})
\label{viterbi2}
\tag{A.5}
\end{equation}

To record the arguments that maximize Eq. \ref{viterbi2}, the array $\xi_t$ is introduced. Viterbi algorithm begins by initializing $\nu_t$ and $\xi_t$ for $t=1$, as shown in Eqs. \ref{viterbi3} and \ref{viterbi4}.

\begin{equation}
\nu_1(i) =  \pi_ib_i(o_1), \;\;\;\;\;\;\; i \in [1, N]
\label{viterbi3}
\tag{A.6}
\end{equation}

\begin{equation}
    \xi_1(i) =  0, \;\;\;\;\;\;\;\;\;\;\;\;\;\;\;\;\; i \in [1, N]
\label{viterbi4}
\tag{A.7}
\end{equation}

The algorithm then recursively solves for each time step $t = 2 \; to \; T$ and each state $j = 1 \; to \; N$, using Eqs. \ref{viterbi5} and \ref{viterbi6}.

\begin{equation}
\nu_t(j) =  \underset{i = 1, ..,N}{\max} \nu_{t-1}(i)a_{ij}b_j(o_t), \;\;\; t \in [2, T], j \in [1,N]
\label{viterbi5}
\tag{A.8}
\end{equation}

\begin{equation}
\xi_t(j) = \underset{i = 1, ..,N}{\argmax} \nu_{t-1}(i)a_{ij}, \;\;\; t \in [2, T], j \in [1,N]
\label{viterbi6}
\tag{A.9}
\end{equation}

Finally, the probability of the best score sequence is determined using Eq. \ref{viterbi7} and the best path state is discovered in Eq. \ref{viterbi8}.

\begin{equation}
P^* =  \underset{i = 1, ..,N}{\max} \nu_T(i)
\label{viterbi7}
\tag{A.10}
\end{equation}

\begin{equation}
q_T^* = \underset{i = 1, ..,N}{\argmax} \;\; \nu_T(i)
\label{viterbi8}
\tag{A.11}
\end{equation}

Eq. \ref{viterbi9} finds out the best state sequence starting at the best path state and backtracking in time by following $\xi_t$.

\begin{equation}
    q_t^* = \xi_{t+1}(q_{t+1}^*), \;\;\; t = T-1, T-2,.. ,1
\label{viterbi9}
\tag{A.12}
\end{equation}

\ifCLASSOPTIONcompsoc
\else
  \section*{Acknowledgment}
\fi


\ifCLASSOPTIONcaptionsoff
  \newpage
\fi




\begin{thebibliography}{2}

\bibitem{fireEye2020}
FireEye M-Trends 2020 report, [online]
\\https://content.fireeye.com/m-trends/rpt-m-trends-2020.

\bibitem{lukaseder}
T. Lukaseder, "Security in high-bandwidth networks," \emph{Open Access Repository of the University of Ulm}, Dissertation, 2020.
\\http://dx.doi.org/10.18725/OPARU-33154

\bibitem{gilder}
G. Gilder, “TELECOSM: How infinite bandwidth will revolutionize our world,” \emph{Free Press}, 2000.

\bibitem{sommer}
R. Sommer, and V. Paxson, "Outside the closed world: On using machine learning for network intrusion detection," in \emph{Proc. IEEE Symp. Secur. Priv.}, May 2010, pp. 305-316.

\bibitem{zeekCluster}
Zeek cluster architecture, [online] 
\\https://docs.zeek.org/en/current/cluster/index.html.

\bibitem{snort}
Snort intrusion detection/prevention system, [online] 
\\https://www.snort.org.

\bibitem{zeek}
Zeek network security monitor, [online]
\\https://www.zeek.org.

\bibitem{mai}
J. Mai, C. Chuah, A. Sridharan, T. Ye, and H. Zang, "Is sampled data sufficient for anomaly detection?," in \emph{Proc. IMC'06: 6th ACM SIGCOMM Conf. Internet Measurement}, Oct. 2006, pp. 165-176.

\bibitem{iscx}
A. Shiravi, H. Shiravi, M. Tavallaee, and A. Ghorbani, "Toward developing a systematic approach to generate benchmark datasets for intrusion detection," \emph{Computers \& Security}, vol. 31, no. 3, pp. 357-374, 2012.

\bibitem{cic}
I. Sharafaldin, A. H. Lashkari, and A. Ghorbani, "Toward generating a new intrusion detection dataset and intrusion traffic characterization," in \emph{Proc. ICISSP 2018}, pp. 108-116, 2018.

\bibitem{jonsson}
E. Jonsson, and T. Olovsson, "A quantitative model of the security intrusion process on attacker behavior," in \emph{IEEE Trans. Softw. Engg.}, vol. 23, no. 4, pp. 235-245, 1997.

\bibitem{mehta}
V. Mehta, C. Bartzis, H. Zhu, E. Clarke, and J. Wing, "Ranking attack graphs," in \emph{Proc. Recent Adv. Intr. Det.}, 2006, pp. 127-144.

\bibitem{sawilla}
R. Sawilla, and X. Ou, "Identifying critical attack assets in dependency attack graphs," in \emph{Proc. Eur. Sym. Res. Comp. Sec.}, 2008, pp. 18-34.

\bibitem{google}
S. Brin, and L. Page, "The anatomy of a large-scale hypertextual Web search engine," in \emph{Compt Net. and ISDN Sys.}, vol. 30, no. 1-7, pp. 107-117, 1998.

\bibitem{HMMRabiner}
L. Rabiner, "A tutorial on hidden markov models and selected applications in speech recognition," in \emph{Proc. of the IEEE}, vol. 77, no. 2, pp. 257-286, 1989.

\bibitem{googleDataFlow}
T. Akidau, R. Bradshaw, C. Chambers, S. Chernyak, R. Fernandez-Moctezuma, R. Lax, S. McVeety, D. Mills, F. Perry, E. Schmidt, and S. Whittle, "The dataflow model: a practical approach to balancing correctness, latency, and cost in massive-scale, unbounded, out-of-order data processing," in \emph{Proc. VLDB Endowment}, vol. 8, no. 12, 2015.

\bibitem{cvss}
Common vulnerability scoring system, [online]
\\https://www.first.org/cvss/.

\bibitem{nessus}
Nessus network vulnerability scanner, [online]
\\https://www.tenable.com/products/nessus/.

\bibitem{openvas}
OpenVAS - open vulnerability assessment scanner, [online]
\\https://www.openvas.org/.

\bibitem{nmap}
Nmap: the network mapper, [online]
\\https://nmap.org/.

\bibitem{srivastava}
U. Srivastava, and J. Widom, "Flexible time management in data stream systems," in \emph{Proc. ACM PODS 2004}, Jun. 2004, pp. 263-274.

\bibitem{mitre}
MITRE ATT\&CK, [online]
\\https://attack.mitre.org/.

\bibitem{splunk}
Splunk enterprise security, [online]
\\https://www.splunk.com/.

\bibitem{extrahop}
ExtraHop, [online]
\\https://www.extrahop.com/.

\bibitem{logrhythm}
LogRhythm, [online]
\\https://logrhythm.com/.

\bibitem{HMMIntro}
D. Jurasky, and J. Martin, "Speech and language processing", \emph{Pearson Prentice Hall}, 2009. 

\bibitem{shawly}
T. Shawly, A. Elghariani, J. Kobes, and, A. Ghafoor, "Architectures for detecting interleaved multi-stage network attacks using hidden Markov models," \emph{IEEE Trans. Dep. Sec. Comp.}, 2019.

\bibitem{duffield}
N. Duffield, C. Lund, and M. Thorup, "Properties and prediction of flow statistics from sampled packet
streams," in \emph{Proc. ACM SIGCOMM IMW}, Nov. 2002.

\bibitem{javed}
Y. Javed, M. Felemban, T. Shawly, J. Kobes, and, A. Ghafoor, "A partition-driven integrated security architecture for cyberphysical systems," in \emph{IEEE Comp.}, vol. 53, no. 3, pp. 47-56, 2020.

\bibitem{zeeshan}
Z. Ahmad, A. Khan, C. Shiang, J. Abdullah, and F. Ahmad, "Network intrusion detection system: A systematic study of machine learning and deep learning approaches," \emph{Trans. Emerg. Tel. Tech.}, vol. 32, no. 1, Art. no. e4150, Oct. 2020.

\bibitem{chaabouni}
N. Chaabouni, M. Mosbah, A. Zemmari, C. Sauvignac, and P. Faruki, "Network intrusion detection for IoT security based on learning techniques," \emph{IEEE Comm. Surv. Tutor.}, vol. 21, no. 3, pp. 2671-2701, 2019.

\bibitem{LiMIT}
J. Li, D. Lim, and K. Sollins,"Dependency-based distributed intrusion detection," in \emph{Proc. DETER community workshop on cyber secur. experimentation and test}, USENIX Association, 2007.

\bibitem{hardegen}
C. Hardegen, M. Petersen, C. Ezelu, T. Geier, S. Rieger, and U. Buehler,"A hierarchical architecture and probabilistic
strategy for collaborative intrusion detection," in \emph{Proc. 2021 IEEE CNS}, Oct. 2021.

\bibitem{bovenzi}
G. Bovenzi, G. Aceto, D. Ciuomzo, V. Persico, and A. Pescape,"A hierarchical hybrid intrusion detection approach in IoT scenarios," in \emph{Proc. 2020 IEEE Globecom}, Dec. 2020.

\bibitem{hamids}
H. R. Ghaeini, and N. O. Tippenhauer,"HAMIDS: Hierarchical monitoring intrusion detection system for industrial control systems," in \emph{CPS-SPC 2016}, Oct. 2016.

\bibitem{xie}
Y. Xie, Y. Wang, H. He, Y. Xiang, S. Yu, and X. Liu, "A general framework for modeling and perceiving distributed network behavior," \emph{IEEE/ACM Trans. on Netw.}, vol. 24, no. 5, pp. 3162-3176, Oct. 2016.

\bibitem{alkadi}
O. Alkadi, N. Moustafa, B. Turnbull, and K. Choo, "A deep blockchain framework-enabled collaborative intrusion detection for protecting IoT and cloud networks," \emph{IEEE Internet of Things Journal}, May 2020.

\bibitem{husak}
M. Husak, J. Komarkova, E. Bou-Harb, and P. Celeda, "Survey of attack projection, prediction and forecasting in cyber security," \emph{IEEE Comm. Surv. and Tutor.}, vol. 21, no. 1, pp. 640-660, 2019.

\bibitem{fu}
C. Fu, Q. Li, M. Shen, and K. Xu, "Realtime robust malicious traffic detection via frequency domain analysis," in \emph{Proc. 2021 ACM SIGSAC Conf. on Comm. and Comp. Security}, Nov. 2021.

\bibitem{mirsky}
Y. Mirsky, T. Doitshman, Y. Elovici, and A. Shabtai, "Kitsune: An ensemble of autoencoders for online network intrusion detection," in \emph{Proc. 2018 ISOC NDSS}, Feb. 2018.

\bibitem{cho}
S. Cho, "Incorporating soft computing techniques into a probabilistic intrusion detection system," \emph{IEEE Trans. Sys. Man. Cyber.}, vol. 32, no. 2, pp. 154-160, 2002.

\bibitem{joshi}
S. Joshi, and V. Phoha, "Investigating hidden Markov models capabilities in anomaly detection," in \emph{Proc. ACM-SE 43: 43rd Annual Southeast Region. Conf.}, Mar. 2005, pp. 98-103.

\bibitem{larrohando}
G. Florez-Larrahondo, S. Bridges, and R. Vaughn, "Efficient modeling of discrete events for anomaly detection using hidden Markov models," in \emph{Proc. ISC'05: 8th Intl. Conf. Info. Secur.}, Sept. 2005, pp. 506-514.

\bibitem{ariu}
D. Ariu, R. Tronci, and G. Giacinto, "HMMPayl: An intrusion detection system based on hidden Markov models," \emph{Comp. \& Secur.}, vol. 30, no. 4, pp. 221-241, 2011.

\bibitem{gornitz}
N. Gornitz, M. Braun, and M. Kloft, "Hidden Markov anomaly detection," in \emph{Proc. 32nd Intl. Conf. Mach. Learn.}, 2015, pp. 1833-1842.

\bibitem{haslum}
K. Haslum, M. Moe, and S. Knapskog, "Real-time intrusion prevention and security analysis of networks using HMMs," in \emph{Proc. 2008 Intl. Conf. Local Comp. Netw.}, 2008. 

\bibitem{farhadi}
H. Farhadi, M. AmirHaeri, and, M. Khansari, "Alert correlation and prediction using data mining and HMM," \emph{ISC Intl. J. Info. Secur.}, vol. 2, no. 2, pp. 77-101, 2011.  

\bibitem{sendi}
A. Sendi, M. Dagenais, M. Jabbarifar, and M. Couture "Real time intrusion prediction based on optimized alerts with hidden Markov model," \emph{J. Netw.}, vol. 7, no. 2, pp. 311-321, 2012.

\bibitem{holgado}
P. Halgodo, V. Villagra, and L. Vazquez, "Real-time multistep attak predeiction based on hidden Markov models," \emph{IEEE Trans. Dep. Sec. Comp.}, vol. 17, no. 1, pp. 134-147, 2017.

\bibitem{chadza}
T. Chadza, K. Kyriakopoulos, and S. Lambotharan, "Learning to learn sequential network attacks using hidden Markov models," \emph{IEEE Access}, vol. 8, pp. 134480-134497, 2020.

\bibitem{androulidakis}
G. Androulidakis, V. Chatzigiannakis, and S. Papavassiliou, "Network anomaly detection and classification via opportunistic sampling," \emph{IEEE Network}, vol. 23, no. 1, pp. 6-12, 2009. 

\bibitem{bartos}
K. Bartos, and M. Rehak, "IFS: Intelligent flow sampling for network security-an adaptive approach," \emph{Intl. J. Netw. Mgmt.}, vol. 25, pp. 263-282, 2015.

\bibitem{su}
L. Su, Y. Yao, N. Li, J. Liu, Z. Lu, and B. Liu, "Hierarchical clustering based network traffic data reduction for improving suspicious flow detection," in \emph{Proc. 12th IEEE Intl. Conf. Big Data Sci. and Engg.}, Sept. 2018.

\bibitem{meng}
W. Meng, "Intrusion detection in the era of IoT: building trust via traffic filtering and sampling," \emph{IEEE Computer}, vol. 51, no. 7, pp. 36-43, 2018. 

\bibitem{sibai}
R. Sibai, Y. Chabchoub, C. Jaoude, J. Demejian, and M. Togbe, "Towards efficient data sampling for temporal anomaly detection in sensor networks," in \emph{Proc. IEEE MENACOMM 2019}, Nov. 2019.

\end{thebibliography}
%

%

\begin{IEEEbiographynophoto}{Yahya Javed}
is a Ph.D. candidate at the Elmore Family School of Electrical and Computer Engineering, Purdue University. His research interests include security analytics, design of intelligent security information and event management systems, and development of resilient enterprise systems by leveraging artificial intelligence-based techniques.
\end{IEEEbiographynophoto}

\begin{IEEEbiographynophoto}{Mosab A. Khayat}
is an Assistant Professor in the Department of Computer Engineering at Umm Al-Qura University, Makkah, Saudi Arabia. His research interests include modeling and evaluation of visual analytics systems, visualization, and machine learning. Khayat received his Ph.D. from the Elmore Family School of Electrical and Computer Engineering, Purdue University.
\end{IEEEbiographynophoto}


\begin{IEEEbiographynophoto}{Ali A. Elghariani}
 [S’12–M’14] received his B.S. and M.S. degrees in electrical and electronic engineering from the University of Tripoli, Tripoli, Libya. He obtained his Ph.D. degree in communications, networking, and signal processing from the Elmore Family School of Electrical and Computer Engineering, Purdue University, in 2014. Currently, he is a research engineer at XCOM-Labs in San Diego, CA. He was a visiting researcher at Purdue University in 2017-2019. He was a Lecturer at the Department of Electrical and Electronic Engineering, University of Tripoli, Tripoli, Libya in 2015-2016. His research interests include network security modeling, signal detection and channel estimation in large-scale MIMO systems, millimeter wave communication and optimization techniques in wireless communications.
\end{IEEEbiographynophoto}

\begin{IEEEbiographynophoto}{Arif Ghafoor}
is currently a professor in the Elmore Family School of Electrical and Computer Engineering, Purdue University. His research interests are in the areas of multimedia and database systems, and cyber security. He has served on the Editorial Boards of several journals.
\end{IEEEbiographynophoto}




\end{document}